%% file: cancun.tex
\documentclass[multphys,vecphys]{svmult}

\usepackage{makeidx}     
\usepackage{graphicx}    
\usepackage{multicol}    

\makeindex             

\begin{document}

\title*{Lectures on the Theory of Cosmological Perturbations}
\titlerunning{Cosmological Perturbations} 

\author{Robert H. Brandenberger}
\institute{Brown University Physics Department, Providence, RI 02912, USA
\texttt{rhb@het.brown.edu}}
%
%
\maketitle

\begin{abstract}

The theory of cosmological perturbations has become a cornerstone
of modern quantitative cosmology since it is the framework which provides the
link between the models of the very early Universe such as the
inflationary Universe scenario (which yield causal mechanisms for
the generation of fluctuations) and the wealth of recent
high-precision data on the spectrum of density fluctuations and
cosmic microwave anisotropies. In these lectures, I provide an
overview of the classical and quantum theory of cosmological
fluctuations. 

Crucial points in both the current inflationary paradigm 
\cite{Guth,Lindebook} of the
early Universe and in proposed alternatives such as the 
Pre-Big-Bang \cite{PBB}
and Ekpyrotic \cite{EKP} scenarios are that, first, the 
perturbations are generated on
microscopic scales as quantum vacuum fluctuations, and, second,
that via an accelerated expansion of the background geometry (or by
a contraction of the background), the wavelengths of the
fluctuations become much larger than the Hubble radius for a long
period of cosmic evolution. Hence, both Quantum Mechanics and General
Relativity are required in order to understand the generation and
evolution of fluctuations.

As a guide to develop the physical intuition for
the evolution of inhomogeneities, I begin with a discussion of the 
Newtonian theory of fluctuations. applicable at late times and on
scales smaller than the Hubble radius. The analysis of super-Hubble
fluctuations requires a general relativistic analysis. I first review
the classical relativistic theory of fluctuations, and then discuss
their quantization. I conclude with a brief overview of two
applications of the theory of cosmological fluctuations: the
trans-Planckian ``problem'' of inflationary cosmology
and the current status of the study of
the back-reaction of cosmological fluctuations on the background
space-time geometry. Most of this article is based on the
review \cite{MFB} to which the reader is referred to for the details
omitted in these lecture notes.  

\end{abstract}

\section{Motivation}

As described in the lectures by Tegmark at this school \cite{Tegmark},
observational cosmology is currently in its golden years. Using a
variety of observational techniques, physicists and astronomers are
exploring the large-scale structure of the Universe. The Cosmic
Microwave Background (CMB) is the observational window which in recent
years has yielded the most information. The anisotropies in the CMB
have now been detected on a wide range of angular scales, giving
us a picture of the Universe at the time of recombination, the time
that the cosmic photons last scattered. Large-scale galaxy
redshift surveys are providing us with increasingly accurate power
spectra of the distribution of objects in the Universe which emit
light, which - modulo the question whether light in fact traces mass
(this is the issue of the cosmic {\it bias}) - gives us the
distribution of mass at the present time. Analyses of the spectra of quasar
absorption line systems and weak gravitational lensing surveys
are beginning to give us complementary information about the
distribution of matter (independent of whether this matter in fact
emits light, thus shedding light on the biasing issue). The analysis of weak 
gravitational lensing maps is in fact sensitive not only to the baryonic 
but also to the dark matter, and promises to give a technique which 
unambiguously reveals where the dark matter is concentrated. 
X-ray telescopes are providing additional information
on the distribution of sources which emit X-rays.

The current data fits astonishingly well with the current paradigm of
early Universe cosmology, the {\it inflationary Universe scenario} 
\cite{Guth}. However, it is important to keep in mind that what is
tested observationally is the paradigm that the primordial spectrum
of inhomogeneities was scale-invariant and predominantly adiabatic
(these terms will be explained in the following Section), and that there
might exist other scenarios of the very early Universe which do not
yield inflation but predict a scale-invariant adiabatic spectrum.
For example, within both the Pre-Big-Bang \cite{PBB} and the
Ekpyrotic scenarios \cite{EKP} there may be models which yield
such a spectrum \footnote{Note, however, that whereas the simplest
inflationary models yield an almost scale-invariant $n = 1$ spectrum of
fluctuations, as discussed in detail in these lectures, this is
not the case for the simplest models of Pre-Big-Bang type nor
for four dimensional descriptions of the Ekpyrotic scenario. In
the case of single field realizations of Pre-Big-Bang cosmology, 
a spectrum with spectral index $n = 4$ emerges \cite{BGGMV}. In
Ekpyrotic cosmology, the value of the index of the final power
spectrum is under active debate. Most studies conclude either
that the spectral index is $n = 3$ \cite{Lyth1,BF,Tsujikawa,Hwang2,TBF},
or that the result is ill-defined because of the singularities at
the bounce \cite{Lyth2,MPPS} (see, however, \cite{KOST2,DV,CDC} for
arguments in support of a final scale-invariant spectrum). See also
\cite{KKL} for criticisms of the basic setup of the Ekpyrotic
scenario.}
One should also not forget that topological defect models of
structure formation (see e.g. \cite{ShellVil,HK,RHBtoprev} for
reviews) naturally yield a scale-invariant spectrum,
however of primordial isocurvature nature and thus no longer compatible
with the latest CMB anisotropy results. 

The theory of cosmological perturbations is what allows us to
connect theories of the very early Universe with the data on
the large-scale structure of the Universe at late times and is
thus of central importance in modern cosmology. The techniques
discussed below are applicable to most scenarios of the very
early Universe. Most specific applications mentioned, however, 
will be within the
context of the inflationary Universe scenario. To understand what
the key requirements for a viable theory of cosmological perturbations
are, recall the basic space-time diagram for inflationary cosmology
(Figure 1):
\begin{figure}
\centering
\includegraphics[height=6cm]{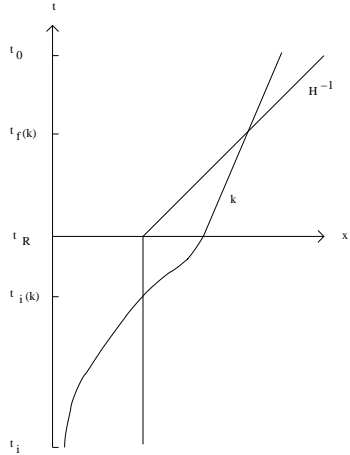}
\caption{Space-time diagram (sketch) showing the evolution
of scales in inflationary cosmology. The vertical axis is
time, and the period of inflation lasts between $t_i$ and
$t_R$, and is followed by the radiation-dominated phase
of standard big bang cosmology. During exponential inflation,
the Hubble radius $H^{-1}$ is constant in physical spatial coordinates
(the horizontal axis), whereas it increases linearly in time
after $t_R$. The physical length corresponding to a fixed
comoving length scale labelled by its wavenumber $k$ increases
exponentially during inflation but increases less fast than
the Hubble radius (namely as $t^{1/2}$), after inflation.}
\label{fig:1}       
\end{figure}
Since, during the phase of standard cosmology
$t_R < t < t_0$, where $t_R$ corresponds to the end of inflation,
and $t_0$ denotes the present time, 
the Hubble radius $l_H(t) \equiv H^{-1}(t)$ expands
faster that the physical wavelength associated with a fixed
comoving scale, the wavelength becomes larger than the Hubble
radius as we go backwards in time. 
However, during the phase of accelerated expansion
(inflation), the physical wavelength increases much faster than
the Hubble radius, and thus at early times the fluctuations
emerged at micro-physical sub-Hubble scales. The idea is
that micro-physical processes (as we shall see, quantum
vacuum fluctuations) are responsible for the origin of the
fluctuations. However, during the period when the wavelength
is super-Hubble, it is essential to describe the fluctuations
using General Relativity. Thus, both Quantum Mechanics and
General Relativity are required to successfully describe the
generation and evolution of cosmological fluctuations.

A similar conclusion can be reached when considering the
space-time diagram in a model of Pre-Big-Bang or Ekpyrotic
type, where the Universe starts out in a contracting phase
during which the Hubble radius contracts faster than the
physical length corresponding to a fixed comoving scale (see
Figure 2).
\begin{figure}
\centering
\includegraphics[height=6cm]{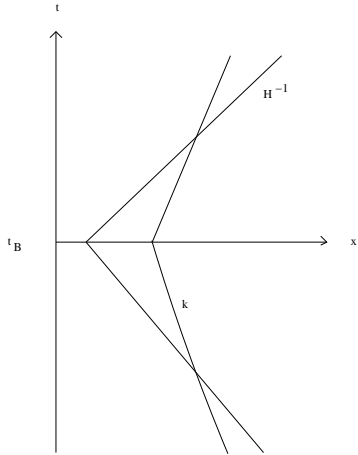}
\caption{Space-time diagram (sketch) showing the evolution
of scales in a cosmology of PBB or Ekpyrotic type. The axes are
as in Figure 1. Times earlier than $t_B$ correspond to the
contracting phase, times after describe the post-bounce phase
of expansion as described in standard cosmology.
The Hubble radius decreases relative to a fixed comoving scale
during the contracting phase, and increases faster in the expanding phase.
Fluctuations of cosmological interest today are generated
sub-Hubble but propagate super-Hubble for a long time interval.}
\label{fig:2}       
\end{figure}
The contracting phase ends at a cosmological bounce, after which
the Universe is assumed to follow the same evolution history as
it does in standard Big Bang cosmology. As in inflationary cosmology,
quantum vacuum fluctuation on sub-Hubble scales (in this case in
the contracting phase) are assumed to be the seeds of the
inhomogeneities observed today. For a long time period, the scale
of the fluctuation is super-Hubble.

Thus, we see that in inflationary cosmology as well as in Pre-Big-Bang
and Ekpyrotic-type models, both Quantum Mechanics and General
Relativity are required to understand the generation and evolution
of cosmological perturbations.

\section{Newtonian Theory of Cosmological Perturbations}
\label{rhbsec:1}

\subsection{Introduction}

The growth of density fluctuations is a consequence of the
purely attractive nature of the gravitational force. Imagine (first
in a non-expanding background)
a density excess $\delta \rho$ localized about some point ${\bf x}$ in space.
This fluctuation produces an attractive force which pulls
the surrounding matter towards ${\bf x}$. The magnitude of this
force is proportional to $\delta \rho$. Hence, by Newton's
second law
\begin{equation} \label{rhbeq1}
\ddot{\delta \rho} \, \sim \, G \delta \rho \, ,
\end{equation}
where $G$ is Newton's gravitational constant. Hence, there is
an exponential instability of flat space-time to the development
of fluctuations. 

Obviously, in General Relativity it is inconsistent to consider
density fluctuations in a non-expanding background. If we
consider density fluctuations in an expanding background,
then the expansion of space leads to a friction term in (\ref{rhbeq1}).
Hence, instead of an exponential instability to the development of
fluctuations, the growth rate of fluctuations in an expanding Universe
will be as a power of time. It is crucial to determine what this power
is and how it depends both on the background cosmological expansion rate
and on the length scale of the fluctuations.

We will be taking the background space-time to be homogeneous
and isotropic, with a metric given by
\begin{equation} \label{background}
ds^2 \, = \, dt^2 - a(t)^2 d{\bf x}^2 \, ,
\end{equation}
where $t$ is physical time, $d{\bf x}^2$ is the Euclidean metric
of the spatial hypersurfaces (here taken for simplicity to be
spatially flat), and $a(t)$ denoting the scale factor, in terms
of which the expansion rate is given by $H(t) = \dot{a} / a $.
The coordinates ${\bf x}$ used above are ``comoving'' coordinates,
coordinates painted onto the expanding spatial hypersurfaces. Note,
however, that in the following two subsections ${\bf x}$ will denote
the physical coordinates, and ${\bf q}$ the comoving ones.

The materials covered in this section are discussed in several
excellent textbooks on cosmology, e.g. in 
\cite{Weinberg,Peebles,Padmanabhan,Peacock}.
 
\subsection{Perturbations about Minkowski Space-Time}

To develop some physical intuition, we first consider the
evolution of hydrodynamical matter fluctuations in a fixed
non-expanding background. Note that in this case the background
Einstein equations are {\bf not} satisfied. 

In this context, matter is described
by a perfect fluid, and gravity by the Newtonian gravitational
potential $\varphi$. The fluid variables are the energy density
$\rho$, the pressure $p$, the fluid velocity ${\bf v}$, and
the entropy density $S$. The basic hydrodynamical equations
are
\begin{eqnarray} \label{rhbeq2}
\dot{\rho} + \nabla \cdot (\rho {\bf v}) & = & 0 \nonumber \\
\dot{{\bf v}} + ({\bf v} \cdot \nabla) {\bf v} + {1 \over {\rho}} \nabla p
+ \nabla \varphi & = & 0 \nonumber \\
\nabla^2 \varphi & = & 4 \pi G \rho \\
\dot{S} + ({\bf v} \cdot \nabla) S & = & 0 \nonumber \\
p & = & p(\rho, S) \, . \nonumber
\end{eqnarray}
The first equation is the continuity equation, the second is the Euler
(force) equation, the third is the Poisson equation of Newtonian gravity,
the fourth expresses entropy conservation, and the last describes
the equation of state of matter. The derivative with respect to time
is denoted by an over-dot.

The background is given by the background energy density $\rho_o$, 
the background
pressure $p_0$, vanishing velocity, constant gravitational potential
$\varphi_0$ and constant entropy density $S_0$. As mentioned above, it
does {\bf not} satisfy the background Poisson equation.

The equations for cosmological perturbations are obtained by perturbing
the fluid variables about the background,
\begin{eqnarray} \label{rhbeq3}
\rho & = & \rho_0 + \delta \rho \nonumber \\
{\bf v} & = & \delta {\bf v} \nonumber \\
p & = & p_0 + \delta p \\
\varphi & = & \varphi_0 + \delta \varphi \nonumber \\
S & = & S_0 + \delta S \, , \nonumber
\end{eqnarray}
where the fluctuating fields $\delta \rho, \delta {\bf v}, \delta p,
\delta \varphi$ and $\delta S$ are functions of space and time, by
inserting these expressions into the basic hydrodynamical equations
(\ref{rhbeq2}), by linearizing, and by combining the resulting equations
which are of first order in time to obtain the following second order
differential equations for the energy density fluctuation $\delta \rho$
and the entropy perturbation $\delta S$
\begin{eqnarray} \label{rhbeq4}
\ddot{\delta \rho} - c_s^2 \nabla^2 \delta \rho - 4 \pi G \rho_0 \delta \rho
& = & \sigma \nabla^2 \delta S \\
\dot \delta S \ & = & 0 \, , \nonumber
\end{eqnarray}
where the variables $c_s^2$ and $\sigma$ describe the equation of state
\begin{equation} \label{pressurepert}
\delta p \, = \, c_s^2 \delta \rho + \sigma \delta S
\end{equation}
with 
\begin{equation}
c_s^2 \, = \, \bigl({{\delta p} \over {\delta \rho}}\bigr)_{|_{S}}
\end{equation}
denoting the square of the speed of sound.

What can we learn from these equations? First of all, since
the equations are linear, we can work in Fourier space. Each
Fourier component $\delta \rho_k(t)$ of the fluctuation field 
$\delta \rho({\bf x}, t)$ 
\begin{equation}
\delta \rho ({\bf x}, t) \, = \, 
\int e^{i {\bf k} \cdot {\bf x}} \delta \rho_k(t)
\end{equation}
evolves independently.

There are various types of fluctuations. If the entropy fluctuation
$\delta S$ vanishes, we have {\bf adiabatic} fluctuations. If
the entropy fluctuation $\delta S$ is non-vanishing but 
$\dot{\delta \rho} = 0$, we speak on an {\bf entropy} fluctuation.

The first conclusions we can draw from the basic perturbation
equations (\ref{rhbeq4}) are that \\
1) entropy fluctuations do not grow, \\
2) adiabatic fluctuations are time-dependent, and \\
3) entropy fluctuations seed an adiabatic mode.

Taking a closer look at the equation of motion (\ref{rhbeq4}) for
$\delta \rho$, we see that the third term on the left hand side
represents the force due to gravity, a purely attractive force 
yielding an instability of flat space-time to the development of
density fluctuations (as discussed earlier, see (\ref{rhbeq1})).
The second term on the left hand side of (\ref{rhbeq4}) represents
a force due to the fluid pressure which tends to set up pressure waves.
In the absence of entropy fluctuations, the evolution of $\delta \rho$
is governed by the combined action of both pressure and gravitational
forces.

Restricting our attention to adiabatic fluctuations, we see from
(\ref{rhbeq4}) that there is a critical wavelength, the Jeans length,
whose wavenumber $k_J$ is given by
\begin{equation} \label{Jeans}
k_J \, = \, \bigl({{4 \pi G \rho_0} \over {c_s^2}}\bigr)^{1/2} \, .
\end{equation}
Fluctuations with wavelength longer than the Jeans length ($k \ll k_J$)
grow exponentially
\begin{equation} \label{expgrowth}
\delta \rho_k(t) \, \sim \, e^{\omega_k t} \,\, {\rm with} \,\,
\omega_k \sim 4 (\pi G \rho_0)^{1/2}
\end{equation}
whereas short wavelength modes ($k \gg k_J$) oscillate with 
frequency $\omega_k \sim c_s k$. Note that the value of the
Jeans length depends on the equation of state of the background.
For a background dominated by relativistic radiation, the Jeans
length is large (of the order of the Hubble radius $H^{-1}(t)$), 
whereas for pressure-less matter the Jeans length
goes to zero.

\subsection{Perturbations about an Expanding Background}

Let us now improve on the previous analysis and study Newtonian
cosmological fluctuations about an expanding background. In this
case, the background equations are consistent (the non-vanishing
average energy density leads to cosmological expansion). However,
we are still neglecting general relativistic effects (the 
fluctuations of the metric). Such effects
turn out to be dominant on length scales larger than the Hubble
radius $H^{-1}(t)$, and thus the analysis of this section is
applicable only to scales smaller than the Hubble radius. 

The background cosmological model is given by the energy density
$\rho_0(t)$, the pressure $p_0(t)$, and the recessional velocity
${\bf v}_0 = H(t) {\bf x}$, where ${\bf x}$ is the Euclidean spatial
coordinate vector (``physical coordinates''). The space- and time-dependent
fluctuating fields are defined as in the previous section:
\begin{eqnarray} \label{fluctansatz2}
\rho(t, {\bf x}) & = & \rho_0(t) \bigl(1 + \delta_{\epsilon}(t, {\bf x}) 
\bigr)\nonumber \\
{\bf v}(t, {\bf x}) & = & {\bf v}_0(t, {\bf x}) + \delta {\bf v}(t, {\bf x}) 
\\
p(t, {\bf x}) & = & p_0(t) + \delta p(t, {\bf x}) \, , \nonumber 
\end{eqnarray}
where $\delta_{\epsilon}$ is the fractional energy density perturbation
(we are interested in the fractional rather than in the absolute energy
density fluctuation!), and the pressure perturbation $\delta p$ is
defined as in (\ref{pressurepert}). In addition, there is the
possibility of a non-vanishing entropy perturbation defined as in
(\ref{rhbeq3}).

We now insert this ansatz into the basic hydrodynamical equations 
(\ref{rhbeq2}), linearize in the perturbation variables, and combine
the first order differential equations 
for $\delta_{\epsilon}$ and $\delta p$ into a single second order
differential equation for $\delta \rho_{\epsilon}$. The result simplifies
if we work in ``comoving coordinates'' ${\bf q}$ which are the coordinates
painted onto the expanding background, i.e. 
\begin{equation}
{\bf x}(t) \, = \, a(t) {\bf q}(t) \, .
\end{equation}
After a substantial amount of algebra, we obtain the following equation
which describes the time evolution of density fluctuations:
\begin{equation} \label{Newtoneq}
\ddot{\delta_{\epsilon}} + 2 H \dot{\delta_{\epsilon}} 
- {{c_s^2} \over {a^2}} \nabla_q^2 \delta_{\epsilon} 
- 4 \pi G \rho_0 \delta_{\epsilon} \, 
= \, {{\sigma} \over {\rho_0 a^2}} \delta S \, ,
\end{equation}
where the subscript $q$ on the $\nabla$ operator indicates that derivatives
with respect to comoving coordinates are used.
In addition, we have the equation of entropy conservation
\begin{equation}
\dot{\delta S} \, = \, 0 \, .
\end{equation}

Comparing with the equations (\ref{rhbeq4}) obtained in the absence of
an expanding background, we see that the only difference is the presence
of a Hubble damping term in the equation for $\delta_{\epsilon}$. This
term will moderate the exponential instability of the background to
long wavelength density fluctuations. In addition, it will lead to a
damping of the oscillating solutions on short wavelengths. More specifically,
for physical wavenumbers $k_p \ll k_J$ (where $k_J$ is again given by
(\ref{Jeans})), and in a matter-dominated background cosmology, the
general solution of (\ref{Newtoneq}) in the absence of any entropy
fluctuations is given by
\begin{equation} \label{Newtonsol}
\delta_k(t) \, = \, c_1 t^{2/3} + c_2 t^{-1} \, ,
\end{equation}
where $c_1$ and $c_2$ are two constants determined by the initial
conditions, and we have dropped the subscript $\epsilon$ in expressions
involving $\delta_{\epsilon}$. 
There are two fundamental solutions, the first is a
growing mode with $\delta_k(t) \sim a(t)$, the second a decaying
mode with $\delta_k(t) \sim t^{-1}$.
On short wavelength, one obtains damped oscillatory motion:
\begin{equation} \label{Newtonsolosc}
\delta_k(t) \, \sim \, a^{-1/2}(t) exp \bigl( \pm i c_s k \int dt' a^{-1}(t')
\bigr) \, .
\end{equation}

As a simple application of the Newtonian equations for cosmological
perturbations derived above, let us compare the predicted cosmic
microwave background (CMB) anisotropies in a spatially
flat Universe with only baryonic matter - Model A -
to the corresponding anisotropies
in a flat Universe with mostly cold dark matter (pressure-less non-baryonic
dark matter) - Model B. We start with the observationally known amplitude
of the relative density fluctuations today (time $t_0$), 
and we use the fact that
the amplitude of the CMB anisotropies on the angular scale $\theta(k)$
corresponding to the comoving wavenumber $k$ is set by the value of
the primordial gravitational potential $\phi$ - introduced in the
following section - which in turn is related to the value of the
primordial density fluctuations at Hubble radius crossing (and {\bf not}
to its value of the time $t_{rec}$). See e.g. Chapter 17 of \cite{MFB}).

In Model A, the dominant component of the pressure-less matter is
coupled to radiation between $t_{eq}$ and $t_{rec}$, the time of
last scattering. Thus, the Jeans length is comparable to the Hubble
radius. Therefore, for comoving galactic scales, $k \gg k_J$ in this
time interval, and thus the fractional density contrast decreases
as $a(t)^{-1/2}$. In contrast, in Model B, the dominant component of
pressure-less matter couples only weakly to radiation, and hence
the Jeans length is negligibly small. Thus, in Model B, the
relative density contrast grows as $a(t)$ between $t_{eq}$ and $t_{rec}$.
In the time interval $t_{rec} < t < t_0$, the fluctuations scale
identically in Models A and B. Summarizing,
we conclude, working backwards in time from a fixed amplitude
of $\delta_k$ today, that the amplitudes of $\delta_k(t_{eq})$ in Models
A and B (and thus their primordial values) are related by
\begin{equation}
\delta_k(t_{eq})|_{A} \, \simeq \, 
\bigl({{a(t_{rec})} \over {a(t_{eq})}} \bigr) \delta_k(t_{eq})|_{B} \, .
\end{equation}
Hence, in Model A (without non-baryonic dark matter) the CMB anisotropies
are predicted to be a factor of about 30 larger 
\cite{SW} than in Model B, way
in excess of the recent observational results. This is one of the
strongest arguments for the existence of non-baryonic dark matter. 

\subsection{Characterizing Perturbations}

Let us consider perturbations on a fixed
comoving length scale given by a comoving wavenumber $k$. 
The corresponding physical length increases
as $a(t)$. This is to be compared to the Hubble radius $H^{-1}(t)$
which scales as $t$ provided $a(t)$ grows as a power of $t$. In
the late time Universe, $a(t) \sim t^{1/2}$ in the radiation-dominated
phase (i.e. for $t < t_{eq}$, and $a(t) \sim t^{2/3}$ in the 
matter-dominated period ($t_{eq} < t < t_0$). 
Thus, we see that at sufficiently early times, all comoving scales
had a physical length larger than the Hubble radius. If we consider 
large cosmological scales (e.g. those corresponding to the observed
CMB anisotropies or to galaxy clusters), the time $t_H(k)$ of 
``Hubble radius crossing'' (when the physical length was equal to the
Hubble radius) was in fact later than $t_{eq}$. As we will see
in later sections, the time of Hubble radius crossing plays an
important role in the evolution of cosmological perturbations.

Cosmological fluctuations can be described either in position space
or in momentum space. In position space, we compute the root mean
square mass fluctuation $\delta M / M(k, t)$ in a sphere of radius
$l = 2 \pi / k$ at time $t$. A scale-invariant spectrum of fluctuations is
defined by the relation
\begin{equation} \label{scaleinv}
{{\delta M} \over M}(k, t_H(k)) \, = \, {\rm const.} \, .
\end{equation}
Such a spectrum was first suggested by Harrison \cite{Harrison}
and Zeldovich \cite{Zeldovich} as a reasonable choice for the
spectrum of cosmological fluctuations. We can introduce the ``spectral
index'' $n$ of cosmological fluctuations by the relation
\begin{equation} \label{specindex}
\bigl({{\delta M} \over M}\bigr)^2(k, t_H(k)) \, \sim \, k^{n - 1} \, ,
\end{equation}
and thus a scale-invariant spectrum corresponds to $n = 1$.

To make the transition to the (more frequently used) momentum space
representation, we Fourier decompose the fractional spatial density
contrast
\begin{equation} \label{Fourier}
\delta_{\epsilon}({\bf x}, t) \, = \, 
\int d^3k {\tilde{\delta_{\epsilon}}}({\bf k}, t) e^{i {\bf k} \cdot {\bf x}}
\, .
\end{equation}
The {\bf power spectrum} $P_{\delta}$ of density fluctuations is defined by
\begin{equation} \label{densspec}
P_{\delta}(k) \, = \, k^3 |{\tilde{\delta_{\epsilon}}}(k)|^2 \, ,
\end{equation}
where $k$ is the magnitude of ${\bf k}$, and we have assumed for simplicity
a Gaussian distribution of fluctuations in which the amplitude of the
fluctuations only depends on $k$.

We can also define the power spectrum of the gravitational potential $\varphi$:
\begin{equation} \label{gravspec}
P_{\varphi}(k) \, = \, k^3 |{\tilde{\delta \varphi}}(k)|^2 \, .
\end{equation}
These two power spectra are related by the Poisson equation (\ref{rhbeq2})
\begin{equation} \label{relspec}
P_{\varphi}(k) \, \sim \, k^{-4} P_{\delta}(k) \, .
\end{equation}

In general, the condition of scale-invariance is expressed in momentum
space in terms of the power spectrum evaluated at a fixed time. To obtain
this condition, we first use the time dependence of the 
fractional density fluctuation from (\ref{Newtonsol}) to determine
the mass fluctuations at a fixed time $t > t_H(k) > t_{eq}$ (the last
inequality is a condition on the scales considered)
\begin{equation} \label{timerel}
\bigl({{\delta M} \over M}\bigr)^2(k, t) \, = \,
\bigl({t \over {t_H(k)}}\bigr)^{4/3} 
\bigl({{\delta M} \over M}\bigr)^2(k, t_H(k)) \, .
\end{equation}
The time of Hubble radius crossing is given by
\begin{equation} \label{Hubble}
a(t_H(k)) k^{-1} \, = \, 2 t_H(k) \, ,
\end{equation} 
and thus
\begin{equation} \label{Hubble2}
t_H(k)^{1/2} \, \sim \, k^{-1} \, .
\end{equation}
Inserting this result into (\ref{timerel}) making use of (\ref{specindex})
we find 
\begin{equation} \label{spec2}
\bigl({{\delta M} \over M}\bigr)^2(k, t) \, \sim \, k^{n + 3} \, .
\end{equation}
Since, for reasonable values of the index of the power spectrum, 
$\delta M / M (k, t)$ is dominated by the Fourier modes with
wavenumber $k$, we find that (\ref{spec2}) implies
\begin{equation} \label{spec3}
|{\tilde{\delta_{\epsilon}}}|^2 \, \sim \, k^{n} \, ,
\end{equation}
or, equivalently,
\begin{equation} \label{spec4}
P_\varphi(k) \, \sim \, k^{n - 1} \, .
\end{equation}
 
\subsection{Matter Fluctuations in the Radiation Era}

Let us now briefly consider fluctuations in the radiation dominated
epoch. We are interested in both the fluctuations in radiation
and in matter (cold dark matter). In the Newtonian treatment,
Eq. (\ref{Newtoneq}) is replaced by separate equations for each
matter fluid component (these components are designated by the
labels A or B):
\begin{equation} \label{Newtoneq2}
\ddot{\delta_A} + 2 H \dot{\delta_A} - v_A^2 a^{-2} \nabla^2 \delta_A \,
= \, 4 \pi G \sum_B \rho_B \delta_B \, ,
\end{equation}
where $\rho_B$ indicate the background densities, and $\delta_B$
the fractional density fluctuations. The velocities of the respective
fluid components are denoted by $v_B$, with $v_r^2 = 1/3$ for radiation
and $v_m = 0$ for cold dark matter.

In the radiation dominated epoch, the evolution of the fluctuations in
radiation is to a first approximation (in the ratio of the background
densities) independent of the cold matter content. Inserting the expansion
rate for this epoch, we thus immediately obtain
\begin{equation}
\delta_r(t) \, \sim \, a(t)^2 
\end{equation}
on scales much larger than the Hubble scale, i.e. $k \ll k_H$, whereas
$\delta_r$ undergoes damped oscillatory motion on smaller scales.

The evolution of the matter fluctuation $\delta_m$ is more complicated.
Its equation of motion is dominated by the source term coming from $\delta_r$.
What results is logarithmic growth of the amplitude of $\delta_m$,
instead of the growth proportional to $a(t)$ which would occur on
these scales in the absence of radiation. This damping effect on matter
fluctuations due to the presence of radiation is called the ``Meszaros
effect''. It leads to a turnover in the spectrum of cosmological
fluctuations at a scale $k_{eq}$ which crosses the Hubble radius at
the time of equal matter and radiation. On larger scales ($k < k_{eq}$),
one has the primordial power spectrum with spectral index $n$, on smaller
scales, to a first approximation, the spectral index changes to $n - 4$.
The details of the power spectrum on small scales depend largely on
the specifics of the matter content in the Universe. One can write
\begin{equation}
P_{final}(k, t) \, = \, T(k, t) P_0(k, t)
\end{equation}
where $P_0$ is the primordial power spectrum extrapolated to late
times with unchanged spectral index, and $P_{final}$ denotes the
actual power spectrum which depends on effects such as the ones mentioned
above. For more details see e.g. \cite{Peebles,Padmanabhan}.
 
\section{Relativistic Theory of Cosmological Fluctuations}

\subsection{Introduction}

The Newtonian theory of cosmological fluctuations discussed in the
previous section breaks down on scales larger than the Hubble radius
because it neglects perturbations of the metric, and because on large
scales the metric fluctuations dominate the dynamics.

Let us begin with a heuristic argument to show why metric fluctuations
are important on scales larger than the Hubble radius. For such
inhomogeneities, one should be able to approximately describe the
evolution of the space-time by applying the first 
Friedmann-Lem\^aitre-Robertson-Walker (FLRW) equation of homogeneous
and isotropic cosmology to the local Universe (this approximation is
made more rigorous in \cite{Afshordi}):
\begin{equation} \label{FRW1}
\bigl({{{\dot a}} \over a}\bigr)^2 \, = \, {{8 \pi G} \over 3} \rho \, .
\end{equation}
Based on this equation, a large-scale fluctuation of the
energy density will lead to a fluctuation (``$\delta a$'') of
the scale factor $a$ which grows in time. This is due to the fact
that self gravity amplifies fluctuations even on length scales $\lambda$
greater than the Hubble radius.

This argument is made rigorous in the following analysis of cosmological
fluctuations in the context of general relativity, where both metric
and matter inhomogeneities are taken into account. We will consider
fluctuations about a homogeneous and isotropic background cosmology,
given by the metric (\ref{background}), which can be written in
conformal time $\eta$ (defined by $dt = a(t) d\eta$) as
\begin{equation} \label{background2}
ds^2 \, = \, a(\eta)^2 \bigl( d\eta^2 - d{\bf x}^2 \bigr) \, .
\end{equation}
The evolution of the scale factor is determined by the two FLRW equations,
(\ref{FRW1}) and
\begin{equation} \label{FRW2}
\dot{\rho} \, = \, - 3 H (\rho + p) \, ,
\end{equation}
which determine the expansion rate and its time derivative in terms
of the equation of state of the matter, whose background stress-energy
tensor can be written as
\begin{equation}
T^{\mu}_{\nu} \, = \, \left(
\begin{array} {cccc}
\rho & 0 & 0 & 0 \\
0 & -p & 0 & 0 \\
0 & 0 & -p & 0 \\
0 & 0 & 0 & -p
\end{array}
\right) \, .
\end{equation}

The theory of cosmological perturbations is based on expanding the Einstein
equations to linear order about the background metric. The theory was
initially developed in pioneering works by Lifshitz \cite{Lifshitz}. 
Significant progress in the understanding of the physics of cosmological
fluctuations was achieved by Bardeen \cite{Bardeen} who realized the
importance of subtracting gauge artifacts (see below) from the
analysis (see also \cite{PV}). The following discussion is based on
Part I of the comprehensive review article \cite{MFB}. Other reviews - in
some cases emphasizing different approaches - are 
\cite{Kodama,Ellis,Hwang,Durrer}.

\subsection{Classifying Fluctuations}

The first step in the analysis of metric fluctuations is to
classify them according to their transformation properties
under spatial rotations. There are scalar, vector and second rank
tensor fluctuations. In linear theory, there is no coupling
between the different fluctuation modes, and hence they evolve
independently (for some subtleties in this classification, see
\cite{Stewart}). 

We begin by expanding the metric about the FLRW background metric
$g_{\mu \nu}^{(0)}$ given by (\ref{background2}):
\begin{equation} \label{pertansatz}
g_{\mu \nu} \, = \, g_{\mu \nu}^{(0)} + \delta g_{\mu \nu} \, .
\end{equation}
The background metric depends only on time, whereas the metric
fluctuations $\delta g_{\mu \nu}$ depend on both space and time.
Since the metric is a symmetric tensor, there are at first sight
10 fluctuating degrees of freedom in $\delta g_{\mu \nu}$.

There are four degrees of freedom which correspond to scalar metric
fluctuations (the only four ways of constructing a metric from
scalar functions):
\begin{equation} \label{scalarfl}
\delta g_{\mu \nu} \, = \, a^2 \left(
\begin{array} {cc}
2 \phi & -B_{,i} \\
-B_{,i} & 2\bigl(\psi \delta_{ij} - E_{,ij} \bigr) 
\end{array}
\right) \, ,
\end{equation}
where the four fluctuating degrees of freedom are denoted (following
the notation of \cite{MFB}) $\phi, B, E$, and $\psi$, a comma denotes
the ordinary partial derivative (if we had included spatial curvature
of the background metric, it would have been the covariant derivative
with respect to the spatial metric), and $\delta_{ij}$ is the Kronecker
symbol.

There are also four vector degrees of freedom of metric fluctuations,
consisting of the four ways of constructing metric fluctuations from
three vectors:
\begin{equation} \label{vectorfl}
\delta g_{\mu \nu} \, = \, a^2 \left(
\begin{array} {cc}
0 & -S_i \\
-S_i & F_{i,j} + F_{j,i} 
\end{array}
\right) \, ,
\end{equation}
where $S_i$ and $F_i$ are two divergence-less vectors (for a vector
with non-vanishing divergence, the divergence contributes to the
scalar gravitational fluctuation modes).

Finally, there are two tensor modes which correspond to the two
polarization states of gravitational waves:
\begin{equation} \label{tensorfl}
\delta g_{\mu \nu} \, = \, -a^2 \left(
\begin{array} {cc}
0 & 0 \\
0 & h_{ij} 
\end{array}
\right) \, ,
\end{equation}
where $h_{ij}$ is trace-free and divergence-less
\begin{equation}
h_i^i \, = \, h_{ij}^j \, = \, 0 \, .
\end{equation}

Gravitational waves do not couple at linear order to the matter
fluctuations. Vector fluctuations decay in an expanding background
cosmology and hence are not usually cosmologically important.
The most important fluctuations, at least in inflationary cosmology,
are the scalar metric fluctuations, the fluctuations which couple
to matter inhomogeneities and which are the relativistic generalization
of the Newtonian perturbations considered in the previous section.

\subsection{Gauge Transformation}

The theory of cosmological perturbations is at first sight complicated
by the issue of gauge invariance (at the final stage, however, we will
see that we can make use of the gauge freedom to substantially simplify
the theory). The coordinates $t, {\bf x}$ of space-time carry no
independent physical meaning. They are just labels to designate points
in the space-time manifold. By performing a small-amplitude 
transformation of the space-time coordinates
(called ``gauge transformation'' in the following), we can easily
introduce ``fictitious'' fluctuations in a homogeneous and isotropic
Universe. These modes are ``gauge artifacts''.

We will in the following take an ``active'' view of gauge transformation.
Let us consider two space-time manifolds, one of them a homogeneous
and isotropic Universe ${\cal M}_0$, the other a physical Universe 
${\cal M}$ with inhomogeneities. A choice of coordinates can be considered
to be a mapping ${\cal D}$ between the manifolds ${\cal M}_0$ and ${\cal M}$.
Let us consider a second mapping ${\tilde{\cal D}}$ which will map the
same point (e.g. the origin of a fixed coordinate system) in ${\cal M}_0$
into different points in ${\cal M}$. Using the inverse of these maps
${\cal D}$ and ${\tilde{\cal D}}$, we can assign two different sets of
coordinates to points in ${\cal M}$. 

Consider now a physical quantity $Q$ (e.g. the Ricci scalar)
on ${\cal M}$, and the corresponding
physical quantity $Q^{(0)}$ on ${\cal M}_0$ Then, in the first coordinate
system given by the mapping ${\cal D}$, the perturbation $\delta Q$ of
$Q$ at the point $p \in {\cal M}$ is defined by
\begin{equation}
\delta Q(p) \, = \, Q(p) - Q^{(0)}\bigl({\cal D}^{-1}(p) \bigr) \, .
\end{equation}
Analogously, in the second coordinate system given by ${\tilde{\cal D}}$,
the perturbation is defined by
\begin{equation}
{\tilde{\delta Q}}(p) \, = \, Q(p) - 
Q^{(0)}\bigl({\tilde{{\cal D}}}^{-1}(p) \bigr) \, .
\end{equation}
The difference
\begin{equation}
\Delta Q(p) \, = \, {\tilde{\delta Q}}(p) - \delta Q(p)
\end{equation}
is obviously a gauge artifact and carries no physical significance.

Some of the metric perturbation degrees of freedom introduced in
the first subsection are gauge artifacts. To isolate these, 
we must study how coordinate transformations act on the metric.
There are four independent gauge degrees of freedom corresponding
to the coordinate transformation
\begin{equation}
x^{\mu} \, \rightarrow \, {\tilde x}^{\mu} = x^{\mu} + \xi^{\mu} \, .
\end{equation}
The zero (time) component $\xi^0$ of $\xi^{\mu}$ leads to a scalar metric
fluctuation. The spatial three vector $\xi^i$ can be decomposed
\begin{equation}
\xi^i \, = \, \xi^i_{tr} + \gamma^{ij} \xi_{,j}
\end{equation}
(where $\gamma^{ij}$ is the spatial background metric)
into a transverse piece $\xi^i_{tr}$ which has two degrees of freedom
which yield vector perturbations, and the second term (given by
the gradient of a scalar $\xi$) which gives
a scalar fluctuation. To summarize this paragraph, there are
two scalar gauge modes given by $\xi^0$ and $\xi$, and two vector
modes given by the transverse three vector $\xi^i_{tr}$. Thus,
there remain two physical scalar and two vector
fluctuation modes. The gravitational waves are gauge-invariant. 

Let us now focus on how the scalar gauge transformations (i.e. the
transformations given by $\xi^0$ and $\xi$) act on the scalar
metric fluctuation variables $\phi, B, E$, and $\psi$. An immediate
calculation yields:
\begin{eqnarray}
{\tilde \phi} \, &=& \, \phi - {{a'} \over a} \xi^0 - (\xi^0)^{'} \nonumber \\
{\tilde B} \, &=& \, B + \xi^0 - \xi^{'} \\
{\tilde E} \, &=& \, E - \xi \nonumber \\
{\tilde \psi} \, &=& \, \psi + {{a'} \over a} \xi^0 \, , \nonumber
\end{eqnarray}
where a prime indicates the derivative with respect to conformal time $\eta$.

There are two approaches to deal with the gauge ambiguities. The first is
to fix a gauge, i.e. to pick conditions on the coordinates which
completely eliminate the gauge freedom, the second is to work with a
basis of gauge-invariant variables.

If one wants to adopt the gauge-fixed approach, there are many
different gauge choices. Note that the often used synchronous gauge
determined by $\delta g^{0 \mu} = 0$ does not totally fix the
gauge. A convenient system which completely fixes the coordinates
is the so-called {\bf longitudinal} or {\bf conformal Newtonian gauge}
defined by $B = E = 0$.

If one prefers a gauge-invariant approach, there are many choices
of gauge-invariant variables. A convenient basis first introduced
by \cite{Bardeen} is the basis $\Phi, \Psi$ given by 
\begin{eqnarray} \label{givar}
\Phi \, &=& \, \phi + {1 \over a} \bigl[ (B - E')a \bigr]^{'} \\
\Psi \, &=& \, \psi - {{a'} \over a} (B - E') \, .
\end{eqnarray}
It is obvious from the above equations that the gauge-invariant
variables $\Phi$ and $\Psi$ coincide with the corresponding
diagonal metric perturbations $\phi$ and $\psi$ in longitudinal
gauge. 

Note that the variables defined above are gauge-invariant only
under linear space-time coordinate transformations. Beyond
linear order, the structure of perturbation theory becomes much
more involved. In fact, one can show \cite{SteWa} that the only
fluctuation variables which are invariant under all coordinate
transformations are perturbations of variables which are constant
in the background space-time.

\subsection{Equation of Motion}

We begin with the Einstein equations
\begin{equation} \label{Einstein}
G_{\mu\nu} \, = \, 8 \pi G T_{\mu\nu} \, , 
\end{equation}
where $G_{\mu\nu}$ is the Einstein tensor associated with the space-time
metric $g_{\mu\nu}$, and $T_{\mu\nu}$ is the energy-momentum tensor of matter,
insert the ansatz for metric and matter perturbed about a FLRW 
background $\bigl(g^{(0)}_{\mu\nu}(\eta) ,\, \varphi^{(0)}(\eta)\bigr)$:
\begin{eqnarray} \label{pertansatz2}
g_{\mu\nu} ({\bf x}, \eta) & = & g^{(0)}_{\mu\nu} (\eta) + \delta g_{\mu\nu}
({\bf x}, \eta) \\
\varphi ({\bf x}, \eta) & = & \varphi_0 (\eta) + \delta \varphi
({\bf x}, \eta) \, , 
\end{eqnarray}
(where we have for simplicity replaced general matter by a scalar
matter field $\varphi$)
and expand to linear order in the fluctuating fields, obtaining the
following equations:
\begin{equation} \label{linein}
\delta G_{\mu\nu} \> = \> 8 \pi G \delta T_{\mu\nu} \, .
\end{equation}
In the above, $\delta g_{\mu\nu}$ is the perturbation in the metric and $\delta
\varphi$ is the fluctuation of the matter field $\varphi$.

Note that the components $\delta G^{\mu}_{\nu}$ and $\delta T^{\mu}_{\nu}$
are not gauge invariant. If we want to use the gauge-invariant approach,
we note \cite{MFB} that it is possible to construct a gauge-invariant tensor 
$\delta G^{(gi) \, \mu}_{\nu}$ via
\begin{eqnarray} \label{givar2}
\delta G^{(gi) \, 0}_{0} \, &\equiv& \, \delta G^{0}_{0}
+ (^{(0)}G^{' \, 0}_0)(B - E') \, \nonumber \\
\delta G^{(gi) \, 0}_{i} \, &\equiv& \, \delta G^{0}_{i}
+ (^{(0)}G^0_i - {1 \over 3} {}^{(0)}G^k_k)(B - E')_{,i} \, \\
\delta G^{(gi) \, i}_{j} \, &\equiv& \, \delta G^{i}_{j}
+ (^{(0)}G^{' \, i}_j)(B - E') \, , \nonumber
\end{eqnarray}
where ${}^(0)G^{\mu}_{\nu}$ denote the background values of the
Einstein tensor.
Analogously, a gauge-invariant linearized stress-energy tensor
$\delta T^{(gi) \, \mu}_{\nu}$
can be defined. In terms of these tensors, the gauge-invariant form of
the equations of motion for linear fluctuations reads
\begin{equation} \label{linefin}
\delta G_{\mu\nu}^{(gi)} \> = \> 8 \pi G \delta T_{\mu\nu}^{(gi)} \, .
\end{equation}
If we insert into this equation the ansatz for the general metric
and matter fluctuations (which depend on the gauge), 
only gauge-invariant combinations of the
fluctuation variables will appear.

In a gauge-fixed approach, one can start with the metric in
longitudinal gauge
\begin{equation} \label{longit}
ds^2 \, = \, a^2 \bigl[(1 + 2 \phi) d\eta^2
- (1 - 2 \psi)\gamma_{ij} dx^i dx^j \bigr] \,
\end{equation}
and insert this ansatz into the general perturbation equations
(\ref{linein}). The shortcut of inserting a restricted
ansatz for the metric into the action and deriving the full set
of variational equations is justified in this case. 

Both approaches yield the following set of equations of motion:
\begin{eqnarray} \label{perteom1}
- 3 {\cal H} \bigl( {\cal H} \Phi + \Psi^{'} \bigr) + \nabla^2 \Psi \,
&=& \, 4 \pi G a^2 \delta T^{(gi) \, 0}_0 \nonumber \\
\bigl( {\cal H} \Phi + \Psi^{'} \bigr)_{, i} \,
&=& 4 \pi G a^2 \delta T^{(gi) \, 0}_i \\
\bigl[ \bigl( 2 {\cal H}^{'} + {\cal H}^2 \bigr) \Phi + {\cal H} \Phi^{'}
+ \Psi^{''} + 2 {\cal H} \Psi^{'} \bigr] \delta^i_j && \nonumber \\
+ {1 \over 2} \nabla^2 D \delta^i_j - {1 \over 2} \gamma^{ik} D_{, kj} \,
&=& - 4 \pi G a^2 \delta T^{(gi) \, i}_j \, , \nonumber
\end{eqnarray}
where $D \equiv \Phi - \Psi$ and ${\cal H} = a'/a$. If we work in
longitudinal gauge, then $\delta T^{(gi) \, i}_j = \delta T^i_j$,
$\Phi = \phi$ and $\Psi = \psi$.

The first conclusion we can draw is that if no anisotropic stress
is present in the matter at linear order in fluctuating fields, i.e.
$\delta T^i_j = 0$ for $i \neq j$, then the two metric fluctuation
variables coincide:
\begin{equation} \label{constr}
\Phi \, = \, \Psi \, .
\end{equation}
This will be the case in most simple cosmological models, e.g. in
theories with matter described by a set of scalar fields with
canonical form of the action, and in the case of a perfect fluid
with no anisotropic stress.

Let us now restrict our attention to the case of matter described
in terms of a single scalar field $\varphi$ with action
\begin{equation} \label{sfact}
S \, = \, \int d^4x \sqrt{-g} \bigl[ {1 \over 2} \varphi^{, \alpha}
\varphi_{, \alpha} - V(\varphi) \bigr]
\end{equation}
(where $g$ denotes the determinant of the metric)
and we expand the matter field as
\begin{equation} \label{mfexp}
\varphi ({\bf x}, \eta) \, = \, \varphi_0(\eta) + \delta \varphi({\bf x}, \eta)
\end{equation}
in terms of background matter $\varphi_0$ and matter fluctuation 
$\delta \varphi({\bf x}, \eta)$, then in longitudinal gauge
(\ref{perteom1}) reduce to the following
set of equations of motion (making use of (\ref{constr}))
\begin{eqnarray} \label{perteom2}
\nabla^2 \phi - 3 {\cal H} \phi^{'} - 
\bigl( {\cal H}^{'} + 2 {\cal H}^2 \bigr) \phi \, &=& \, 
4 \pi G \bigl( \varphi^{'}_0 \delta \varphi^{'} + 
V^{'} a^2 \delta \varphi \bigr) \nonumber \\
\phi^{'} + {\cal H} \phi \, &=& \, 4 \pi G \varphi^{'}_0 \delta \varphi \\
\phi^{''} + 3 {\cal H} \phi^{'} + 
\bigl( {\cal H}^{'} + 2 {\cal H}^2 \bigr) \phi \, &=& \, 
4 \pi G \bigl( \varphi^{'}_0 \delta \varphi^{'} - 
V^{'} a^2 \delta \varphi \bigr) \, , \nonumber
\end{eqnarray}
where $V^{'}$ denotes the derivative of $V$ with respect to $\varphi$.
These equations can be combined to give the following second order
differential equation for the relativistic potential $\phi$:
\begin{equation} \label{finaleom}
\phi^{''} + 2 \left( {\cal H} - 
{{\varphi^{''}_0} \over {\varphi^{'}_0}} \right) \phi^{'} - \nabla^2 \phi
+ 2 \left( {\cal H}^{'} - 
{\cal H}{{\varphi^{''}_0} \over {\varphi^{'}_0}} \right) \phi \, = \, 0 \, .
\end{equation}

Let us now discuss this final result for the classical evolution of
cosmological fluctuations. First of all, we note the similarities with
the equation (\ref{Newtoneq}) obtained in the Newtonian theory.
The final term in (\ref{finaleom}) is the force due to gravity leading
to the instability, the second to last term is the pressure force
leading to oscillations (relativistic since we are considering matter
to be a relativistic field), and the second term is the Hubble friction
term. For each wavenumber there are two fundamental solutions. On
small scales ($k > H$), the solutions correspond to damped oscillations,
on large scales ($k < H$) the oscillations freeze out and the dynamics
is governed by the gravitational force competing with the Hubble friction
term. Note, in particular, how the Hubble radius naturally emerges as
the scale where the nature of the fluctuating modes changes from oscillatory
to frozen.

Considering the equation in a bit more detail, observe that if the
equation of state of the background is independent of time (which will be
the case if ${\cal H}^{'} = \varphi^{''}_0 = 0$), that then in an
expanding background, the dominant mode of (\ref{finaleom}) is constant,
and the sub-dominant mode decays. If the equation of state is not constant,
then the dominant mode is not constant in time. Specifically, at the
end of inflation ${\cal H}^{'} < 0$, and this leads to a growth of 
$\phi$ (see the following subsection).

To study the quantitative implications of the equation of motion
(\ref{finaleom}), it is convenient to introduce \cite{BST,BK}
the variable $\zeta$ (which, up to correction terms of the order
$\nabla^2 \phi$ which are unimportant for large-scale fluctuations
is equal to the curvature perturbation ${\cal R}$ in comoving gauge
\cite{Lyth})
by
\begin{equation} \label{zetaeq}
\zeta \, \equiv \, \phi + {2 \over 3} 
{{\bigl(H^{-1} {\dot \phi} + \phi \bigr)} \over { 1 + w}} \, ,
\end{equation}
where
\begin{equation} \label{wvar}
w = {p \over {\rho}}
\end{equation}
characterizes the equation of state of matter. In terms of $\zeta$,
the equation of motion (\ref{finaleom}) takes on the form
\begin{equation}
{3 \over 2} {\dot \zeta} H (1 + w) \, = \, 0 + {\cal O}(\nabla^2 \phi) \, .
\end{equation}
On large scales, the right hand side of the equation is negligible,
which leads to the conclusion that large-scale cosmological fluctuations
satisfy
\begin{equation} \label{zetacons}
{\dot \zeta} (1 + w) \, = \, 0 .
\end{equation}
This implies that except possibly if $1 + w = 0$ at some points in time
during cosmological evolution (which occurs during reheating in inflationary
cosmology if the inflaton field undergoes oscillations - see 
\cite{Fabio1} and \cite{BV,Fabio2} for discussions of the consequences
in single and double field inflationary models, respectively) $\zeta$
is constant. In single matter field models it is indeed possible
to show that ${\dot \zeta} = 0$ on super-Hubble scales independent
of assumptions on the equation of state \cite{Weinberg2,Zhang}.
This ``conservation law'' makes it easy to relate
initial fluctuations to final fluctuations in inflationary cosmology,
as will be illustrated in the following subsection.

\subsection{Application to Inflationary Cosmology}
\label{Sec1}

Let us now return to the space-time sketch of the evolution of
fluctuations in inflationary cosmology (Figure 1) and use the
conservation law (\ref{zetacons}) - in the form
$\zeta = {\rm const}$ on large scales - to relate the amplitude
of $\phi$ at initial Hubble radius crossing during the inflationary
phase (at $t = t_i(k)$) with the amplitude at final Hubble radius
crossing at late times (at $t = t_f(k)$). Since both at early
times and at late times ${\dot \phi} = 0$ on super-Hubble scales
as the equation of state is not changing, (\ref{zetacons})
leads to
\begin{equation} \label{inflcons}
\phi(t_f(k)) \, \simeq \, {{(1 + w)(t_f(k))} \over {(1 + w)(t_i(k))}}
\phi(t_i(k)) \, .
\end{equation} 

This equation will allow us to evaluate the amplitude of the
cosmological perturbations when they re-enter the Hubble radius
at time $t_f(k)$, under the assumption (discussed in detail in
the following section) that the origin of the primordial
fluctuations is quantum vacuum oscillations. 

The time-time perturbed Einstein equation (the first equation
of (\ref{perteom1})) relates the value of $\phi$ at initial
Hubble radius crossing to the amplitude of the relative energy
density fluctuations. This, together with the fact that
the amplitude of the scalar matter field quantum vacuum fluctuations
is of the order $H$, yields
\begin{equation} \label{phiin}
\phi(t_i(k)) \, \sim \, H {{V^{'}} \over V}(t_i(k)) \, .
\end{equation}
In the late time radiation dominated phase, $w = 1/3$,
whereas during slow-roll inflation
\begin{equation} \label{win}
1 + w(t_i(k)) \, \simeq \, {{{\dot \varphi_0}^2} \over V}(t_i(k)) \, .
\end{equation}
Making, in addition, use of the slow roll conditions satisfied
during the inflationary period
\begin{eqnarray} \label{srcond}
H {\dot \varphi_0} \, &\simeq& \, - V^{'}  \nonumber \\
H^2 \, &\simeq& \, {{8 \pi G} \over 3} V \, ,
\end{eqnarray}
we arrive at the final result
\begin{equation} \label{final}
\phi(t_f(k)) \, \sim \, {{V^{3/2}} \over {V^{'}}}(t_i(k)) \, ,
\end{equation}
which gives the position space amplitude of cosmological
fluctuations on a scale labelled by the comoving wavenumber $k$
at the time when the scale re-enters the Hubble radius at
late times, a result first obtained in the case of the
Starobinsky model \cite{Starob3} of inflation in \cite{Mukh1},
and later in the context of scalar field-driven inflation
in \cite{GuthPi,Starob4,Hawking,BST}.

In the case of slow roll inflation, the right hand side of
(\ref{final}) is, to a first approximation, independent of $k$,
and hence the resulting spectrum of fluctuations is
scale-invariant.
 
\section{Quantum Theory of Cosmological Fluctuations}

\subsection{Overview}

As already mentioned in the last subsection of the previous
section, in many models of the very early Universe, in particular
in inflationary cosmology, but also in the Pre-Big-Bang and
in the Ekpyrotic scenarios, primordial inhomogeneities emerge
from quantum vacuum fluctuations on microscopic scales (wavelengths
smaller than the Hubble radius). The wavelength is then stretched
relative to the Hubble radius, becomes larger than the Hubble
radius at some time and then propagates on super-Hubble scales until
re-entering at late cosmological times. In the context of a Universe
with a de Sitter phase, the quantum origin of cosmological fluctuations
was first discussed in \cite{Mukh1} - see \cite{Lukash} for
a more general discussion of the quantum origin of fluctuations
in cosmology, and also \cite{Press,Sato} for
earlier ideas. In particular, Mukhanov \cite{Mukh1} and Press 
\cite{Press} realized that
in an exponentially expanding background, the curvature fluctuations
would be scale-invariant, and Mukhanov provided a quantitative
calculation which also yielded the logarithmic deviation from
exact scale-invariance. 

To understand the role
of the Hubble radius, consider the equation of a free scalar matter field
$\varphi$ on an unperturbed expanding background:
\begin{equation}
\ddot{\varphi} + 3 H \dot{\varphi} - {{\nabla^2} \over {a^2}} \varphi
\, = \, 0 \, .
\end{equation}
The second term on the left hand side of this equation leads to damping
of $\varphi$ with a characteristic decay rate given by $H$. As a
consequence, in the absence of the spatial gradient term, $\dot{\varphi}$
would be of the order of magnitude $H \varphi$. Thus, comparing the
second and the third term on the left hand side, we immediately
see that the microscopic (spatial gradient) term dominates on length
scales smaller than the Hubble radius, leading to oscillatory motion,
whereas this term is negligible on scales larger than the Hubble radius,
and the evolution of $\varphi$ is determined primarily by gravity.

To understand the generation and evolution of fluctuations in current
models of the very early Universe, we thus need both Quantum Mechanics
and General Relativity, i.e. quantum gravity. At first sight, we
are thus faced with an intractable problem, since the theory of quantum
gravity is not yet established. We are saved by the fact that today
on large cosmological scales the fractional amplitude of the fluctuations
is smaller than 1. Since gravity is a purely attractive force, the
fluctuations had to have been - at least in the context of an eternally
expanding background cosmology - very small in the early Universe. Thus,
a linearized analysis of the fluctuations (about a classical
cosmological background) is self-consistent.

From the classical theory of cosmological perturbations discussed in the
previous section, we know that the analysis of scalar metric inhomogeneities
can be reduced - after extracting gauge artifacts -
to the study of the evolution of a single fluctuating
variable. Thus, we conclude that the quantum theory of cosmological
perturbations must be reducible to the quantum theory of a single
free scalar field which we will denote by $v$. 
Since the background in which this scalar field
evolves is time-dependent, the mass of $v$ will be time-dependent. The
time-dependence of the mass will lead to quantum particle production
over time if we start the evolution in the vacuum state for $v$. As
we will see, this quantum particle production corresponds to the
development and growth of the cosmological fluctuations. Thus,
the quantum theory of cosmological fluctuations provides a consistent
framework to study both the generation and the evolution of metric
perturbations. The following analysis is based on Part II of \cite{MFB}.
 
\subsection{Outline of the Analysis}

In order to obtain the action for linearized cosmological
perturbations, we expand the action to quadratic order in
the fluctuating degrees of freedom. The linear terms cancel
because the background is taken to satisfy the background
equations of motion.

We begin with the Einstein-Hilbert action for gravity and the
action of a scalar matter field (for the more complicated
case of general hydrodynamical fluctuations the reader is
referred to \cite{MFB}) 
\begin{equation} \label{action}
S \, = \,  \int d^4x \sqrt{-g} \bigl[ - {1 \over {16 \pi G}} R
+ {1 \over 2} \partial_{\mu} \varphi \partial^{\mu} \varphi - V(\varphi)
\bigr] \, ,
\end{equation}
where $g$ is the determinant of the metric.

The simplest way to proceed is to work in a fixed gauge,
longitudinal gauge, in which the metric and matter take the form
\begin{eqnarray} \label{long}
ds^2 \, &=& \, a^2(\eta)\bigl[(1 + 2 \phi(\eta, {\bf x}))d\eta^2
- (1 - 2 \psi(t, {\bf x})) d{\bf x}^2 \bigr]  \\
\varphi(\eta, {\bf x}) \, 
&=& \, \varphi_0(\eta) + \delta \varphi(\eta, {\bf x}) \nonumber \, .
\end{eqnarray}

The next step is to reduce the number of degrees of freedom. First,
as already mentioned in the previous section, the off-diagonal
spatial Einstein equations force $\psi = \phi$ since
$\delta T^i_j = 0$ for scalar field matter (no anisotropic stresses
to linear order). The two remaining fluctuating variables
$\phi$ and $\varphi$ must be linked by the Einstein constraint
equations since there cannot be matter fluctuations without induced
metric fluctuations. 

The two nontrivial tasks of the lengthy \cite{MFB} computation 
of the quadratic piece of the action is to find
out what combination of $\varphi$ and $\phi$ gives the variable $v$
in terms of which the action has canonical form, and what the form
of the time-dependent mass is. This calculation involves inserting
the ansatz (\ref{long}) into the action (\ref{action}),
expanding the result to second order in the fluctuating fields, making
use of the background and of the constraint equations, and dropping
total derivative terms from the action. In the context of
scalar field matter, the quantum theory of cosmological
fluctuations was developed by Mukhanov \cite{Mukh2,Mukh3} (see
also \cite{Sasaki}). The result is the following
contribution $S^{(2)}$ to the action quadratic in the
perturbations:
\begin{equation} \label{pertact}
S^{(2)} \, = \, {1 \over 2} \int d^4x \bigl[v'^2 - v_{,i} v_{,i} + 
{{z''} \over z} v^2 \bigr] \, ,
\end{equation}
where the canonical variable $v$ (the ``Mukhanov variable'' introduced
in \cite{Mukh3} - see also \cite{Lukash}) is given by
\begin{equation} \label{Mukhvar}
v \, = \, a \bigl[ \delta \varphi + {{\varphi_0^{'}} \over {\cal H}} \phi
\bigr] \, ,
\end{equation}
with ${\cal H} = a' / a$, and where
\begin{equation} \label{zvar}
z \, = \, {{a \varphi_0^{'}} \over {\cal H}} \, .
\end{equation}
In both the cases of power law inflation and slow roll inflation, 
${\cal H}$ and $\varphi_0^{'}$ are proportional and hence
\begin{equation} \label{zaprop}
z(\eta) \, \sim \, a(\eta) \, .
\end{equation}
Note that the variable $v$ is related to the curvature
perturbation ${\cal R}$ in comoving coordinates introduced
in \cite{Lyth} and closely related to the variable $\zeta$ used
in \cite{BST,BK}:
\begin{equation} \label{Rvar}
v \, = \, z {\cal R} \, .
\end{equation}

The equation of motion which follows from the action (\ref{pertact}) is
\begin{equation} \label{pertEOM}
v^{''} - \nabla^2 v - {{z^{''}} \over z} v \, = \, 0 \, ,
\end{equation}
or, in momentum space
\begin{equation} \label{pertEOM2}
v_k^{''} + k^2 v_k - {{z^{''}} \over z} v_k \, = \, 0 \, ,
\end{equation}
where $v_k$ is the k'th Fourier mode of $v$. As a consequence of
(\ref{zaprop}), the mass term in the above equation is given
by the Hubble scale
\begin{equation}
k_H^2 \, \equiv \, {{z^{''}} \over z} \, \simeq \, H^2 \, .
\end{equation}
Thus, it immediately follows from (\ref{pertEOM2}) that on small
length scales, i.e. for
$k > k_H$, the solutions for $v_k$ are constant amplitude oscillations . 
These oscillations freeze out at Hubble radius crossing,
i.e. when $k = k_H$. On longer scales ($k \ll k_H$), the solutions
for $v_k$ increase as $z$:
\begin{equation} \label{squeezing}
v_k \, \sim \, z \,\, , \,\,\,   k \ll k_H \, .
\end{equation}

Given the action (\ref{pertact}), the quantization of the cosmological
perturbations can be performed by canonical quantization (in the same
way that a scalar matter field on a fixed cosmological background
is quantized \cite{BD}). 

The final step in the quantum theory of cosmological perturbations
is to specify an initial state. Since in inflationary cosmology,
all pre-existing classical fluctuations are red-shifted by the
accelerated expansion of space, one usually assumes (we will
return to a criticism of this point when discussing the
trans-Planckian problem of inflationary cosmology) that the field
$v$ starts out at the initial time $t_i$ mode by mode in its vacuum
state. Two questions immediately emerge: what is the initial time $t_i$,
and which of the many possible vacuum states should be chosen. It is
usually assumed that since the fluctuations only oscillate on sub-Hubble
scales, that the choice of the initial time is not important, as long
as it is earlier than the time when scales of cosmological interest
today cross the Hubble radius during the inflationary phase. The
state is usually taken to be the Bunch-Davies vacuum (see e.g. \cite{BD}),
since this state is empty of particles at $t_i$ in the coordinate frame
determined by the FLRW coordinates (see e.g. \cite{RB84} for a
discussion of this point), and since the Bunch-Davies state is
a local attractor in the space of initial states in an expanding
background (see e.g. \cite{BHill}). Thus, we choose the initial
conditions
\begin{eqnarray} \label{incond}
v_k(\eta_i) \, = \, {1 \over {\sqrt{2 \omega_k}}} \\
v_k^{'}(\eta_i) \, = \, {{\sqrt{\omega_k}} \over {\sqrt{2}}} \, \, \nonumber
\end{eqnarray} 
where here $\omega_k = k$, and $\eta_i$ is the conformal time corresponding
to the physical time $t_i$.

Let us briefly summarize the quantum theory of cosmological perturbations.
In the linearized theory, fluctuations are set up at some initial
time $t_i$ mode by mode in their vacuum state. While the wavelength
is smaller than the Hubble radius, the state undergoes quantum
vacuum fluctuations. The accelerated expansion of the
background redshifts the length scale beyond the Hubble radius. The 
fluctuations freeze out when the length scale is equal to the Hubble
radius. On larger scales, the amplitude of $v_k$ increases as the
scale factor. This corresponds to the squeezing of the quantum state present
at Hubble radius crossing (in terms of classical general relativity, it
is self-gravity which leads to this growth of fluctuations). As
discussed e.g. in \cite{PolStar}, the squeezing of the quantum vacuum state
leads to the emergence of the classical nature of the fluctuations.

\subsection{Application to Inflationary Cosmology}

In this subsection we will use the quantum theory of cosmological
perturbations developed in this section to calculate the spectrum
of curvature fluctuations in inflationary cosmology. 

We need to compute the power spectrum ${\cal P}_{\cal R}(k)$ 
of the curvature fluctuation ${\cal R}$ defined in (\ref{Rvar}), namely
\begin{equation}
{\cal R} \, = \, z^{-1} v \, = \, 
\phi + \delta \varphi {{\cal H} \over {\varphi_0^{'}}}
\end{equation}
The idea in calculating the power spectrum at a late time $t$ is
to first relate the power spectrum via the growth rate (\ref{squeezing})
of $v$ on super-Hubble scales to the power spectrum at the time $t_H(k)$
of Hubble radius crossing, and to then use the constancy of the amplitude
of $v$ on sub-Hubble scales to relate it to the initial conditions
(\ref{incond}). Thus
\begin{eqnarray} \label{finalspec1}
{\cal P}_{\cal R}(k, t) \, \equiv  \, k^3 {\cal R}_k^2(t) \, 
&=& \, k^3 z^{-2}(t) |v_k(t)|^2 \\
&=& \, k^3 z^{-2}(t) \bigl( {{z(t)} \over {z(t_H(k))}} \bigr)^2
|v_k(t_H(k))|^2 \nonumber \\
&=& \, k^3 z^{-2}(t_H(k)) |v_k(t_H(k))|^2 \nonumber \\
&\sim& \, k^3 a^{-2}(t_H(k)) |v_k(t_i)|^2 \, , \nonumber
\end{eqnarray}
where in the final step we have used (\ref{zaprop}) and the
constancy of the amplitude of $v$ on sub-Hubble scales. Making use
of the condition 
\begin{equation} \label{Hubble3}
a^{-1}(t_H(k)) k \, = \, H 
\end{equation}
for Hubble radius crossing, and of the
initial conditions (\ref{incond}), we immediately see that
\begin{equation} \label{finalspec2}
{\cal P}_{\cal R}(k, t) \, \sim \, k^3 k^{-2} k^{-1} H^2 \, ,
\end{equation}
and that thus a scale invariant power spectrum with amplitude
proportional to $H^2$ results, in agreement with what was
argued on heuristic grounds in Section (\ref{Sec1}).

\subsection{Quantum Theory of Gravitational Waves}

The quantization of gravitational waves parallels the
quantization of scalar metric fluctuations, but is
more simple because there are no gauge ambiguities. Note
again that at the level of linear fluctuations, scalar metric
fluctuations and gravitational waves are independent. Both
can be quantized on the same cosmological background determined
by background scale factor and background matter. However, in
contrast to the case of scalar metric fluctuations, the tensor
modes are also present in pure gravity (i.e. in the absence of
matter).

Starting point is the action (\ref{action}). Into this
action we insert the metric which corresponds to a 
classical cosmological background plus tensor metric
fluctuations:
\begin{equation}
ds^2 \, = \, a^2(\eta) \bigl[ d\eta^2 - (\delta_{ij} + h_{ij}) dx^i dx^j 
\bigr]\, ,
\end{equation}
where the second rank tensor $h_{ij}(\eta, {\bf x})$ represents the
gravitational waves, and in turn can be decomposed as
\begin{equation}
h_{ij}(\eta, {\bf x}) \, = \, h_{+}(\eta, {\bf x}) e^+_{ij}
+ h_{x}(\eta, {\bf x}) e^x_{ij}
\end{equation}
into the two polarization states. Here, $e^{+}_{ij}$ and $e^{x}_{ij}$ are
two fixed polarization tensors, and $h_{+}$ and $h_{x}$ are the two 
coefficient functions.

To quadratic order in the fluctuating fields, the action separates into
separate terms involving $h_{+}$ and $h_{x}$. Each term is of the form
\begin{equation} \label{actgrav}
S^{(2)} \, = \, \int d^4x {{a^2} \over 2} \bigl[ h'^2 - (\nabla h)^2 \bigr] 
\, ,
\end{equation}
leading to the equation of motion
\begin{equation}
h_k^{''} + 2 {{a'} \over a} h_k^{'} + k^2 h_k \, = \, 0 \, .
\end{equation}
The variable in terms of which the action (\ref{actgrav}) has canonical
kinetic term is
\begin{equation} \label{murel}
\mu_k \, \equiv \, a h_k \, ,
\end{equation}
and its equation of motion is
\begin{equation}
\mu_k^{''} + \bigl( k^2 - {{a''} \over a} \bigr) \mu_k \, = \, 0 \, .
\end{equation}
This equation is very similar to the corresponding equation (\ref{pertEOM2}) 
for scalar gravitational inhomogeneities, except that in the mass term
the scale factor $a(\eta)$ is replaced by $z(\eta)$, which leads to a
very different evolution of scalar and tensor modes during the reheating
phase in inflationary cosmology during which the equation of state of the
background matter changes dramatically.
 
Based on the above discussion we have the following theory for the
generation and evolution of gravitational waves in an accelerating
Universe (first developed by Grishchuk \cite{Grishchuk}): 
waves exit as quantum vacuum fluctuations at the initial time
on all scales. They oscillate until the length scale crosses the Hubble
radius. At that point, the oscillations freeze out and the quantum state
of gravitational waves begins to be squeezed in the sense that
\begin{equation}
\mu_k(\eta) \, \sim \, a(\eta) \, ,
\end{equation}
which, from (\ref{murel}) corresponds to constant amplitude of $h_k$.
The squeezing of the vacuum state leads to the emergence of classical
properties of this state, as in the case of scalar metric fluctuations.

\section{The Trans-Planckian Window}

Whereas the contents of the previous sections are well established,
this and the following section deal with aspects of the theory of
cosmological perturbations which are currently under investigation
and are at the present time rather controversial. First, we consider
the trans-Planckian issue (this section is adapted from \cite{RHBtaiwan}).

The same background dynamics which yields the causal generation mechanism
for cosmological fluctuations, the most spectacular success of inflationary
cosmology, bears in it the nucleus of the ``trans-Planckian problem''. This
can be seen from Fig. 3. If inflation lasts only slightly
longer than the minimal time it needs to last in order to solve the
horizon problem and to provide a
causal generation mechanism for CMB fluctuations, then the 
corresponding physical wavelength of these fluctuations
is smaller than the Planck length at the beginning of the period of inflation.
The theory of cosmological perturbations is based on classical general
relativity coupled to a weakly coupled scalar field description of
matter. Both the theories of gravity and of matter will break down
on trans-Planckian scales, and this immediately leads to the trans-Planckian
problem: are the predictions of standard inflationary cosmology robust
against effects of trans-Planckian physics \cite{RHBrev}?

\begin{figure}
\centering
\includegraphics[height=8cm]{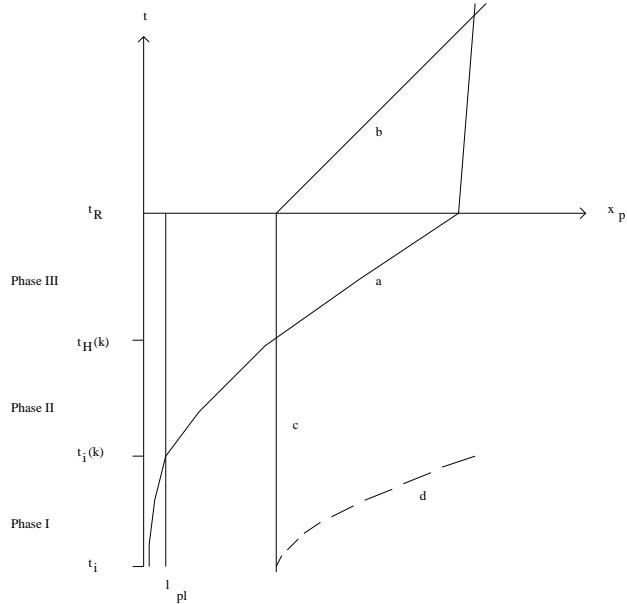}
\caption{Space-time diagram (physical distance vs. time)
showing the origin of
the trans-Planckian problem of inflationary cosmology: at very
early times, the wavelength is smaller than the
Planck scale $\ell _{\rm Pl}$ (Phase I), at intermediate times
it is larger than $\ell _{\rm Pl}$ but smaller than the Hubble
radius $H^{-1}$ (Phase II), and at late times during
inflation it is larger than the Hubble radius (Phase III).
The line labeled a) is the physical wavelength associated
with a fixed comoving scale $k$. The line b) is the Hubble radius
or horizon in SBB cosmology. Curve c) shows the
Hubble radius during inflation. The horizon in inflationary
cosmology is shown in curve d).
\label{FigTP}}
\end{figure}

The simplest way of modeling the possible effects of trans-Planckian
physics, while keeping the mathematical analysis simple, is to replace
the linear dispersion relation $\omega _{_{\rm phys}}=k_{\rm phys}$
of the usual equation for cosmological perturbations by
a non standard dispersion relation $\omega _{_{\rm phys}}=\omega
_{_{\bf phys}}(k)$ which differs from the standard one only
for physical wavenumbers larger than the Planck scale. This method
was introduced \cite{Unruh,CJ} in the context of studying the dependence of
the thermal spectrum of black hole radiation on trans-Planckian physics. 
In the context of cosmology, it has been shown \cite{MB,BM,Niemeyer} 
that this amounts to replacing $k^2$ appearing in
(\ref{pertEOM2}) with $k_{\rm eff}^2(n,\eta )$ defined by
\begin{equation}
k^2 \, \rightarrow \, k_{\rm eff}^2(k,\eta ) \equiv 
a^2(\eta )\omega _{_{\rm phys}}^2\biggl(\frac{k}{a(\eta )}\biggr).
\end{equation}
For a fixed comoving mode, this implies that the dispersion relation
becomes time-dependent. Therefore, the equation of motion of the
quantity $v_k(\eta)$ takes the form (with $z(\eta) \propto a(\eta)$)
\begin{equation} 
\label{eom2}
v_k'' + \biggl[k_{\rm eff}^2(k,\eta ) - {{a''} 
\over a}\biggr]v_k \, = \, 0 \, .
\end{equation}
A more rigorous derivation of this 
equation, based on a variational principle, has been provided \cite{LLMU} 
(see also Ref.~\cite{jacobson}).

The evolution of modes thus must be considered separately in three
phases, see Fig. 3. In Phase I the wavelength is smaller
than the Planck scale, and trans-Planckian physics can play
an important role. In Phase II, the wavelength is larger than the
Planck scale but smaller than the Hubble radius. In this phase,
trans-Planckian physics will have a negligible effect
(this statement can be quantified \cite{Shenker}). Hence,
by the analysis of the previous section, the wave function of fluctuations is
oscillating in this phase, 
\begin{equation}
\label{vsubH}
v_k \, = \, B_1\exp(-ik\eta )+B_2\exp(ik\eta )
\end{equation}
with constant coefficients $B_1$ and $B_2$. In the standard approach,
the initial conditions are fixed in this region and the usual choice
of the vacuum state leads to $B_1=1/\sqrt{2k}$, $B_2=0$.  
Phase III starts at the time $t_{\rm H}(k)$ when the
mode crosses the Hubble radius. During this phase, the wave function is
squeezed.

One source of trans-Planckian effects \cite{MB,BM} on observations
is the possible non-adiabatic evolution of the wave function
during Phase I. If this occurs, then it is possible that the
wave function of the fluctuation mode is not in its vacuum state when
it enters Phase II and, as a consequence, the coefficients $B_1$ and
$B_2$ are no longer given by the standard expressions above. In this
case, the wave function will not be in its vacuum state when it crosses
the Hubble radius, and the final spectrum will be different. In
general, $B_1$ and $B_2$ are determined by the matching conditions
between Phase I and II. 
If the dynamics is adiabatic
throughout (in particular if the $a''/a$ term is negligible), the WKB
approximation holds and the solution is always given by
\begin{equation} 
\label{WKBsol}
v_k (\eta )\, \simeq \, \frac{1}{\sqrt{2k_{\rm eff}(k,\eta )}}
\exp\biggl(-i\int _{\eta _{\rm i}}^{\eta }k_{\rm eff}{\rm d}\tau \biggr)
\, ,
\end{equation} 
where $\eta_i$ is some initial time. Therefore, if we start with
a positive frequency solution only and use this solution, we find
that no negative frequency solution appears. Deep in Region II where
$k_{\rm eff} \simeq k$ the solution becomes
\begin{equation}
v_k(\eta ) \simeq {1 \over {\sqrt{2k}}} \exp(-i \phi - i k \eta),
\end{equation}
i.e. the standard vacuum solution times a phase which will disappear
when we calculate the modulus. To obtain 
a modification of the inflationary spectrum, it is sufficient to 
find a dispersion relation such that the WKB approximation breaks down 
in Phase I.

A concrete class of dispersion relations
for which the WKB approximation breaks down is
\begin{equation}
\label{disprel}
k_{\rm eff}^2(k,\eta ) = k^2 - k^2 \vert b_m \vert
\biggl[{{\ell_{_{\rm pl}}} \over {\lambda(\eta)}} \biggr]^{2m}, 
\end{equation}
where $\lambda (\eta )=2\pi a(\eta )/k$ is the wavelength of a
mode. If we follow the evolution of the modes in Phases I, II and
III, matching the mode functions and their derivatives at the
junction times, the calculation \cite{MB,BM,MB2}
demonstrates that the final spectral index is modified
and that superimposed oscillations appear. 

However, the above example
suffers from several problems. First, in inflationary models with a
long period of inflationary expansion, 
the dispersion relation (\ref{disprel}) leads to complex
frequencies at the beginning of inflation for scales which are of current 
interest in cosmology. Furthermore, the initial conditions for the
Fourier modes of the fluctuation field have to be set in a region
where the evolution is non-adiabatic and the use of the usual vacuum
prescription can be questioned. These problems can be avoided in a toy
model in which we follow the evolution of fluctuations in a bouncing
cosmological background which is asymptotically flat in the past
and in the future. The analysis \cite{MB3} shows that even in this
case the final spectrum of fluctuations depends on the specific 
dispersion relation used.

An example of a dispersion relation which breaks the WKB approximation in
the trans-Planckian regime but does not lead to the problems mentioned in
the previous paragraph was investigated in \cite{LLMU}. It is a
dispersion relation which is linear for both small and large wavenumbers, but
has an intermediate interval during which the frequency decreases as
the wavenumber increases, much like what happens in (\ref{disprel}).
The violation of the WKB condition occurs for wavenumbers near the local 
minimum of the $\omega(k)$ curve.

A justified criticism against the method summarized in the previous analysis
is that the non-standard dispersion relations used are completely ad hoc,
without a clear basis in trans-Planckian physics. There has been a lot
of recent work \cite{EG1,KN,EG2,Mangano} on the implication of space-space
uncertainty relations \cite{Ven,Gross} on the evolution of fluctuations.
The application of the uncertainty relations on the fluctuations lead to
two effects \cite{Kempf,Hassan}. 
Firstly, the equation of motion of the fluctuations
in modified. Secondly, for fixed comoving length scale $k$, the uncertainty
relation is saturated before a critical time $t_i(k)$. Thus, in addition
to a modification of the evolution, trans-Planckian physics leads to a
modification of the boundary condition for the fluctuation modes. The
upshot of this work is that the spectrum of fluctuations is modified.

In \cite{Ho}, the implications of the stringy
space-time uncertainty relation \cite{Yoneya,MY} 
\begin{equation}
\Delta x_{\rm phys} \Delta t \, \geq \, l_s^2 
\end{equation}
on the spectrum of cosmological fluctuations was
studied. Again, application of this
uncertainty relation to the fluctuations leads to two effects. Firstly,
the coupling between the background and the fluctuations is nonlocal in
time, thus leading to a modified dynamical equation of motion
(a similar modification also results \cite{Sera} from quantum
deformations, another example of a consequence of non-commutative basic
physics). Secondly,
the uncertainty relation is saturated at the time $t_i(k)$ when the physical
wavelength equals the string scale $l_s$. Before that time it does not
make sense to talk about fluctuations on that scale. By continuity,
it makes sense to assume that fluctuations on scale $k$ are created at
time $t_i(k)$ in the local vacuum state (the instantaneous WKB vacuum
state).

Let us for the moment neglect the nonlocal coupling
between background and fluctuation, and thus consider the usual
equation of motion for fluctuations in an accelerating background
cosmology.  We distinguish two ranges of scales. 
Ultraviolet modes are generated at late times when the Hubble radius is
larger than $l_s$. On these scales, the spectrum of fluctuations does not
differ from what is predicted by the standard theory, since at the time
of Hubble radius crossing the fluctuation mode will be in its vacuum
state. However, the evolution of infrared modes which are created when
the Hubble radius is smaller than $l_s$ is different. The fluctuations
undergo {\it less} squeezing than they do in the absence of the
uncertainty relation, and hence the final amplitude of fluctuations is
lower. From the equation (\ref{finalspec1}) for the power spectrum of
fluctuations, and making use of the condition
\begin{equation}
a(t_i(k)) \, = \, k l_s 
\end{equation}
for the time $t_i(k)$ when the mode is generated, it follows immediately
that the power spectrum is scale-invariant
\begin{equation} \label{finalb}
{\cal P}_{\cal R}(k) \, \sim \, k^0 \, . 
\end{equation}
In the standard scenario
of power-law inflation the spectrum is red 
(${\cal P}_{\cal R}(k) \sim k^{n-1}$
with $n < 1$). Taking into account the
effects of the nonlocal coupling between background and fluctuation
mode leads \cite{Ho} to a modification of this result: the spectrum
of fluctuations in a power-law inflationary background is in fact blue 
($n > 1$). 

Note that, if we neglect the nonlocal coupling between background and
fluctuation mode, the result of (\ref{finalb}) also holds in a 
cosmological background which is NOT accelerating. Thus, we have a method
of obtaining a scale-invariant spectrum of fluctuations without inflation.
This result has also been obtained in \cite{Wald}, however without a
micro-physical basis for the prescription for the initial conditions.

A key problem with the method of modified dispersion relations
is the issue of back-reaction \cite{Tanaka,Starob}. If the mode
occupation numbers of the fluctuations at Hubble radius crossing are
significant, the danger arises that the back-reaction of the fluctuations
will in fact prevent inflation. Another constraint arises
from the observational limits on the flux of ultra-high-energy
cosmic rays. Such cosmic rays would be produced \cite{Tkachev} in
the present Universe if Trans-Planckian effects of the type
discussed in this section were present. 

An approach to the trans-Planckian issue pioneered by 
Danielsson \cite{Dan} which has recently received a lot of
attention is to avoid the issue of the unknown trans-Planckian
physics and to start the evolution of the fluctuation modes at
the mode-dependent time when the wavelength equals the limiting
scale. Obviously, the resulting spectrum will depend sensitively
on which state is taken to be the initial state. The vacuum
state is not unambiguous, and the choice of a state minimizing
the energy density depends on the space-time splitting \cite{Bozza}.
The signatures of this prescription are typically oscillations
superimposed on the usual spectrum. The amplitude of this effect
depends sensitively on the prescription of the initial state,
and for a fixed prescription also on the background cosmology.
For a discussion of these issues and a list of references on
this approach the reader is referred to \cite{Martin03}.

In summary, due to the exponential red-shifting of wavelengths, 
present cosmological
scales originate at wavelengths smaller than the Planck length early on
during the period of inflation. Thus, Planck physics may well encode
information in these modes which can now be observed in the spectrum of
microwave anisotropies. Two examples have been shown to demonstrate the
existence of this ``window of opportunity'' to probe trans-Planckian
physics in cosmological observations. The first method makes use of
modified dispersion relations to probe the robustness of the predictions
of inflationary cosmology, the second applies the stringy space-time
uncertainty relation on the fluctuation modes. Both methods yield the
result that trans-Planckian physics may lead to measurable effects in
cosmological observables. An important issue which must be studied more
carefully is the back-reaction of the cosmological fluctuations (see e.g.
\cite{ABM} for a possible formalism).

\section{Back-Reaction of Cosmological Fluctuations}

The presence of cosmological fluctuations influences the background
cosmology in which the perturbations evolve. This back-reaction
arises as a second order effect in the cosmological perturbation
expansion. The effect is cumulative in the sense that all
fluctuation modes contribute to the change in the background geometry,
and as a consequence the back-reaction effect can be large even if
the amplitude of the fluctuation spectrum is small. In this
section (based on the review \cite{RHBparis}) we discuss two 
approaches used to quantify back-reaction. In the first approach
\cite{MAB96,ABM},
the effect of the fluctuations on the background is expressed in
terms of an effective energy-momentum tensor. We show that in the
context of an inflationary background cosmology, the long
wavelength contributions to the effective
energy-momentum tensor take the form of a negative cosmological
constant, whose absolute value increases as a function of time since
the phase space of infrared modes is increasing. This then leads to
the speculation \cite{RB98,RB99}
that gravitational back-reaction may lead to a
dynamical cancellation mechanism for a bare cosmological constant,
and yield a scaling fixed point in the asymptotic future in
which the remnant cosmological constant satisfies $\Omega_{\Lambda} 
\sim 1$. We then discuss \cite{GG02}
how infrared modes effect local observables
(as opposed to mathematical background quantities) and find that the
leading infrared back-reaction contributions cancel in single field
inflationary models. However, we expect non-trivial back-reaction of
infrared modes in models with more than one matter field.  

It is well known that gravitational waves
propagating in some background space-time affect the dynamics of
the background. This back-reaction can be described in terms of an 
effective energy-momentum tensor $\tau_{\mu \nu}$. In the short wave limit, 
when the typical wavelength of the waves is small compared with the curvature 
of the background space-time, $\tau_{\mu \nu}$ has the form of a radiative 
fluid with an equation of state $p= \rho / 3$ (where $p$ and $\rho$ denote 
pressure and energy density, respectively). As we have seen in
previous section, in inflationary cosmology it is the long wavelength
scalar metric fluctuations which are more important. Like short
wavelength gravitational waves, these cosmological fluctuations
will contribute to the effective energy-momentum tensor $\tau_{\mu \nu}$.
The work of \cite{MAB96,ABM}
is closely related to work by Woodard and Tsamis \cite{WT1,WT2}
who considered the back-reaction of long wavelength gravitational waves
in pure gravity with a bare cosmological constant. The recent paper
\cite{GG02} is related to the work of Abramo
and Woodard \cite{Abramo1,Abramo2} who initiated the study of back-reaction of
infrared modes on local observables. 

We first review the derivation of the effective
energy-momentum tensor $\tau_{\mu \nu}$ which describes the back-reaction
of linear cosmological fluctuations on the background cosmology,
and summarize the evaluation of this tensor in an inflationary cosmological
background.
This gravitational back-reaction calculation
is related to the early work on the back-reaction of gravitational waves 
by Brill, Hartle and Isaacson \cite{Brill}, 
among others. The idea is to expand the Einstein equations to second order 
in the perturbations, to assume that the first order terms satisfy the 
equations of motion for linearized cosmological perturbations discussed
in previous section 
(hence these terms cancel), to take the spatial average of the remaining 
terms, and to regard the resulting equations as equations for a new 
homogeneous metric $g_{\mu \nu}^{(0, br)}$ which includes the effect of 
the perturbations to quadratic order:
\begin{equation} \label{breq}
G_{\mu \nu}(g_{\alpha \beta}^{(0, br)}) \, = \, 
8 \pi G \left[ T_{\mu \nu}^{(0)} + \tau_{\mu \nu} \right]\, \,
\end{equation}
where the effective energy-momentum tensor $\tau_{\mu \nu}$ of 
gravitational back-reaction contains the terms resulting from spatial 
averaging of the second order metric and matter perturbations:
\begin{equation} \label{efftmunu}
\tau_{\mu \nu} \, = \, < T_{\mu \nu}^{(2)} - 
{1 \over {8 \pi G}} G_{\mu \nu}^{(2)} > \, ,
\end{equation}
where pointed brackets stand for spatial averaging, and the superscripts 
indicate the order in perturbation theory.

As analyzed in detail in \cite{MAB96,ABM}, the back-reaction equation
(\ref{breq}) is covariant under linear space-time coordinate transformations
even though $\tau_{\mu \nu}$ is not invariant \footnote{See \cite{Unruh2}, 
however, for important questions concerning the covariance of the analysis 
under higher order coordinate transformations.}. In the following, we will
work in longitudinal gauge. 

For simplicity, we shall take matter to be described in terms of a single 
scalar field.
By expanding the Einstein and matter energy-momentum tensors to 
second order in the metric and matter fluctuations $\phi$ and 
$\delta \varphi$, respectively, it can be shown that the non-vanishing 
components of the effective back-reaction energy-momentum tensor 
$\tau_{\mu \nu}$ become
\begin{eqnarray}  \label{tzero}
\tau_{0 0} &=& \frac{1}{8 \pi G} \left[ + 12 H \langle \phi \dot{\phi} \rangle
- 3 \langle (\dot{\phi})^2 \rangle + 9 a^{-2} \langle (\nabla \phi)^2
\rangle \right]  \nonumber \\
&+& {1 \over 2} \langle ({\delta\dot{\varphi}})^2 \rangle + {1 \over 2} a^{-2} \langle
(\nabla\delta\varphi)^2 \rangle  \nonumber \\
&+& {1 \over 2} V''(\varphi_0) \langle \delta\varphi^2 \rangle + 2
V'(\varphi_0) \langle \phi \delta\varphi \rangle \quad ,
\end{eqnarray}
and 
\begin{eqnarray}  \label{tij}
\tau_{i j} &=& a^2 \delta_{ij} \left\{ \frac{1}{8 \pi G} \left[ (24 H^2 + 16 
\dot{H}) \langle \phi^2 \rangle + 24 H \langle \dot{\phi}\phi \rangle
\right. \right.  \nonumber \\
&+& \left. \langle (\dot{\phi})^2 \rangle + 4 \langle \phi\ddot{\phi}\rangle
- \frac{4}{3} a^{-2}\langle (\nabla\phi)^2 \rangle \right] + 4 \dot{{%
\varphi_0}}^2 \langle \phi^2 \rangle  \nonumber \\
&+& {1 \over 2} \langle ({\delta\dot{\varphi}})^2 \rangle - {1 \over 6} a^{-2} \langle
(\nabla\delta\varphi)^2 \rangle - 
4 \dot{\varphi_0} \langle \delta \dot{\varphi}\phi \rangle  \nonumber \\
&-& \left. {1 \over 2} \, V''(\varphi_0) \langle \delta\varphi^2
\rangle + 2 V'( \varphi_0 ) \langle \phi \delta\varphi \rangle
\right\} \quad ,
\end{eqnarray}
where $H$ is the Hubble expansion rate.

The metric and matter fluctuation variables $\phi$ and $\delta \varphi$ are 
linked via the Einstein constraint equations, and hence all terms in the 
above formulas for the components of $\tau_{\mu \nu}$ can be expressed in 
terms of two point functions of $\phi$ and its derivatives. The two point 
functions, in turn, are obtained by integrating over all of the Fourier 
modes of $\phi$, e.g.
\begin{equation} \label{tpf}
\langle \phi^2 \rangle \, \sim \,  \int_{k_i}^{k_u} {dk} k^2 \vert \phi_k
\vert^2 \, ,
\end{equation} 
where $\phi_k$ denotes the amplitude of the k'th Fourier mode. The above
expression is divergent both in the infrared and in the ultraviolet. The
ultraviolet divergence is the usual divergence of a free quantum field
theory and can be ``cured" by introducing an ultraviolet cutoff $k_u$. In the
infrared, we will discard all modes $k < k_i$ with wavelength larger than 
the Hubble
radius at the beginning of inflation, since these modes are determined by
the pre-inflationary physics. We take these modes to contribute to the 
background.

At any time $t$ we can separate the integral in (\ref{tpf}) into the 
contribution of infrared and ultraviolet modes, the separation being 
defined by setting the physical wavelength equal to the Hubble radius. 
Thus, in an inflationary Universe the infrared phase space is continually 
increasing since comoving modes are stretched beyond the Hubble radius, 
while the ultraviolet
phase space is either constant (if the ultraviolet cutoff corresponds to a 
fixed physical wavelength), or decreasing (if the ultraviolet cutoff 
corresponds to fixed comoving wavelength). In either case, unless the 
spectrum of the initial fluctuations is extremely blue, two point 
functions such as (\ref{tpf}) will at later stages of an inflationary 
Universe be completely dominated by the infrared sector. In the following, 
we will therefore restrict our attention to this sector, i.e. to wavelengths 
larger than the Hubble radius.

In order to evaluate the two point functions which enter into the expressions 
for $\tau_{\mu \nu}$, we make use of the known time evolution of the linear 
fluctuations $\phi_k$ discussed in previous section. 
On scales larger than the Hubble radius,
and for a time-independent equation of state, $\phi_k$ is constant in time.
From the Einstein constraint equations relating the metric and matter 
fluctuations, and making use of the inflationary slow roll approximation
conditions we find 
\begin{equation} 
\delta \varphi =-{\frac{2V}{V^{\prime }}}\,\phi \,.  \label{constr2}
\end{equation}
Hence, in the expressions (\ref{tzero}) and (\ref{tij}) for $\tau_{\mu \nu}$, 
all terms with space and time derivatives can be neglected, and we obtain
\begin{equation} 
\rho _{br}\equiv \tau _0^0\cong \left( 2\,{\frac{{V^{\prime \prime }V^2}}{{%
V^{\prime }{}^2}}}-4V\right) <\phi ^2>  \label{tzerolong}
\end{equation}
and 
\begin{equation}
p_{br}\equiv -\frac 13\tau _i^i\cong -\rho_{br} \,,  \label{tijlong}
\end{equation}

The main result which emerges from this analysis is that the equation of state
of the dominant infrared contribution to the energy-momentum tensor 
$\tau_{\mu \nu}$ which describes back-reaction takes the form of a 
{\it negative cosmological constant} 
\begin{equation} \label{result}
p_{br}=-\rho _{br} \,\,\, {\rm with} \,\,\, \rho_{br} < 0 \, .
\end{equation}
The second crucial result is that the magnitude of $\rho_{br}$ increases as
a function of time. This is due in part to the fact that, in an inflationary 
Universe, as time increases more and more wavelengths become longer than the 
Hubble radius and begin to contribute to $\rho_{br}$. 

How large is the magnitude of back-reaction? The basic point is that
since the amplitude of each fluctuation mode is small, we need a very
large phase space of infrared modes in order to induce any interesting
effects. In models with a very short period of primordial inflation,
the back-reaction of long-wavelength cosmological fluctuations hence 
will not be important. However, in many single field models
of inflation, in particular in those of chaotic inflation type \cite{Linde},   
inflation lasts so long that the infrared back-reaction effects can build up
to become important for the cosmological background dynamics. 

To give an example, consider chaotic inflation with a potential
\begin{equation} 
V(\varphi )={\frac 12}m^2\varphi ^2\, .  \label{pot}
\end{equation}
In this case, the values of $\phi_k$ for long wavelength modes are well 
known (see e.g. \cite{MFB}), and the integral in (\ref{tpf}) can be
easily performed, thus yielding explicit expressions for the dominant terms
in the effective energy-momentum tensor. Comparing
the resulting back-reaction energy density $\rho_{br}$ 
with the background density $\rho_0$, we find
\begin{equation}   \label{result2}
{\frac{{\rho_{br}(t)} }{{\rho_0}}} \, \simeq \, {\frac{{3} }{{4 \pi}}} {\frac{{
m^2 \varphi_0^2(t_i)} }{{M_P^4}}} \left[ {\frac{{\varphi_0 (t_i)} }{{\varphi_0
(t)}}} \right]^4 \, ,
\end{equation}
where $M_P$ denotes the Planck mass.
Without back-reaction, inflation would end \cite{Linde} when 
$\varphi_0 (t) \, \sim \, M_P$. Inserting this value into (\ref{result2}), 
we see that if  
$\varphi_0 (t_i) \, > \, \varphi_{br} \, \sim \, m^{-1/3} M_P^{4/3}$,
then back-reaction will become important before the end of inflation and 
may shorten the period of inflation. It is interesting to compare this 
value with the scale 
$ \varphi_0 (t_i) \, \sim \, \varphi_{sr} \, = \, m^{-1/2} M_P^{3/2}$
above which the stochastic terms in the scalar field equation of motion
arising in the context of the stochastic approach to chaotic 
inflation \cite{Starob2,Slava} are dominant. Notice that 
since $\varphi_{sr} \gg \varphi_{br}$
(recall that $m \ll M_P$), back-reaction effects can be very important in
the entire range of field values relevant to stochastic inflation.

Since the back-reaction of cosmological fluctuations in an inflationary 
cosmology acts (see (\ref{result})) like a negative cosmological constant,
and since the magnitude of the back-reaction effect increases in time, one
may speculate \cite{RB98} that back-reaction will lead to a dynamical 
relaxation of the cosmological constant (see Tsamis \& Woodard \cite{WT1} 
for similar speculations based on
the back-reaction of long wavelength gravitational waves).

The background metric $g_{\mu \nu}^{(0, br)}$ including back-reaction evolves 
as if the cosmological constant at time $t$ were
\begin{equation} \label{effcosm}
\Lambda_{\rm eff}(t) \, = \, \Lambda_0 + 8 \pi G \rho_{br}(t)
\end{equation}
and not the bare cosmological constant $\Lambda_0$. Hence one might hope to 
identify (\ref{effcosm}) with a time dependent effective cosmological 
constant. Since $\vert \rho_{br}(t) \vert$ increases as $t$ grows, the 
effective cosmological constant will decay. Note that even if the initial 
magnitude  of the perturbations is small, eventually (if inflation lasts a 
sufficiently long time) the back-reaction effect will become large enough to 
cancel any bare cosmological constant.

Furthermore, one might speculate that this dynamical relaxation mechanism for 
$\Lambda$ will be self-regulating. As long as 
$\Lambda_{\rm eff}(t) \, > \, 8 \pi G \rho_m(t)$, where $\rho_m(t)$ stands 
for the energy density in ordinary matter and radiation, the evolution of 
$g_{\mu \nu}^{(0, br)}$ is dominated by $\Lambda_{\rm eff}(t)$. Hence, the 
Universe will be undergoing accelerated expansion, more scales will be 
leaving the Hubble radius and the magnitude of the back-reaction term will 
increase. However, once $\Lambda_{\rm eff}(t)$ falls below $\rho_m(t)$, the 
background will start to decelerate, scales will enter the Hubble radius, 
and the number of modes contributing to the back-reaction will decrease, 
thus reducing the strength of back-reaction. Hence, it is likely that there 
will be a scaling solution to the effective equation of motion for 
$\Lambda_{\rm eff}(t)$ of the form
\begin{equation} \label{scaling}
\Lambda_{\rm eff}(t) \, \sim \, 8 \pi G \rho_m(t) \,.
\end{equation}
Such a scaling solution would correspond to a contribution to the 
relative closure density of $\Omega_{\Lambda} \sim 1$.

There are important concerns about the above formalism, and even more so
about the resulting speculations (many of these were first discussed in
print in \cite{Unruh2}). On a formal level, since our back-reaction effect
is of second order in cosmological perturbation theory, it is necessary
to demonstrate covariance of the proposed back-reaction equation 
(\ref{breq}) beyond linear order, and this has not been done. Next, it might
be argued that by causality super-Hubble fluctuations cannot affect local
observables. Thirdly, from an observational perspective one is not
interested in the effect of fluctuations on the background metric (since
what the background is cannot be determined precisely using local 
observations). Instead, one should compute the back-reaction of cosmological
fluctuations on observables describing the local Hubble expansion rate. One
might then argue that even if long-wavelength fluctuations have an effect
on the background metric, they do not influence local observables.
Finally, it is clear that the speculations in the previous section involve
the extrapolation of perturbative physics deep into the non-perturbative
regime. 

These important issues have now begun to be addressed. Good physical
arguments can be given \cite{Abramo1,Abramo2} supporting the idea that
long-wavelength fluctuations can effect local physics. Consider, for
example, a black hole of mass $M$ absorbing a particle of mass $m$. Even
after this particle has disappeared beyond the horizon, its gravitational
effects (in terms of the increased mass of the black hole) remain
measurable to an external observer. A similar argument can be given in
inflationary cosmology: consider an initial localized mass fluctuation
with a characteristic physical length scale $\lambda$
in an exponentially expanding background. Even after the length scale
of the fluctuation redshifts to be larger than the Hubble radius, the
gravitational potential associated with this fluctuation remains measurable.
On a more technical level, it has recently been shown that super-Hubble
scale (but sub-horizon-scale) metric fluctuations can be parametrically
amplified during inflationary reheating 
\cite{Bassett1,Fabio1,BV,Fabio2}. 
This clearly demonstrates a coupling between local physics and
super-Hubble-scale fluctuations.

These arguments, however, make it even more important to focus on
back-reaction effects of cosmological fluctuations on local physical
observables rather than on the mathematical background metric. 
In recent work \cite{GG02},
the leading infrared back-reaction effects on a local observable
measuring the Hubble expansion rate were calculated.

Consider a perfect fluid with velocity four vector $u^{\alpha}$
in an inhomogeneous cosmological geometry, then the local expansion rate which
generalizes the Hubble expansion rate $H(t)$ of homogeneous isotropic
Friedmann-Robertson-Walker cosmology is given by ${1 \over 3} \Theta$,
where $\Theta$ is the four divergence of $u^{\alpha}$:
\begin{equation}
\Theta \, = \, u^{\alpha}_{; \alpha} \, ,
\end{equation}
the semicolon indicating the covariant derivative.
In \cite{GG02}, the effects of cosmological fluctuations on this variable
were computed to second order in perturbation theory. To leading order
in the infrared expansion, the result is
\begin{equation} \label{brresult1}
\Theta \, = \, 3 {{a^{\prime}} \over {a^2}}
\bigl( 1 - \phi + {3 \over 2} \phi^2 \bigr) - 3 {{\phi^{\prime}} \over a} \, ,
\end{equation}
where the prime denotes the derivative with respect to conformal time.
If we now calculate the spatial average of $\Theta$, the term linear in
$\phi$ vanishes, and - as expected - we are left with a quadratic back-reaction
contribution.

Superficially, it appears from (\ref{brresult1}) that there is a 
non-vanishing back-reaction effect at quadratic order which is not
suppressed for super-Hubble modes. However, we must be careful and
evaluate $\Theta$ not at a constant value of the background coordinates,
but rather at a fixed value of some physical observable. For example,
if we work out the value of $\Theta$ in the case of a matter-dominated
Universe, and express the result as a function of the proper time
$\tau$ given by
$d \tau^2 \, = \, a(\eta)^2 (1 + 2 \phi) d \eta^2$
instead of as a function of conformal time $\eta$, then we find that the
leading infrared terms proportional to $\phi^2$ exactly cancel, and that
thus there is no un-suppressed infrared back-reaction on the local
measure of the Hubble expansion rate.

A more relevant example with respect to the discussion in earlier sections
is a model in which matter is given by a single scalar field. In this
case, the leading infrared back-reaction terms in $\Theta$ are again given by
(\ref{brresult1})
which looks different from the background value $3 H$. However, once again
it is important to express $\Theta$ in terms of a physical background variable.
If we choose the value of the matter field $\varphi$ as this variable,
we find after easy manipulations that, including only the leading
infrared back-reaction terms, 
\begin{equation}
\Theta ( \varphi ) \, = \, \sqrt{3} \sqrt{V(\varphi)} \, . 
\end{equation}
Hence, once again the leading infrared back-reaction contributions vanish,
as already found in the work of \cite{Abramo2} which considered the
leading infrared back-reaction effects on a local observable different than
the one we have used, and applied very different methods \footnote{For a
different approach which also leads to the conclusion that there can be
no back-reaction effects from infrared modes on local observables in
models with a single matter component see \cite{Afshordi}.}.

However, in a model with two matter fields, it is clear that if we
e.g. use the second matter field as a physical clock, then the leading
infrared back-reaction terms will not cancel in $\Theta$, and that thus
in such models infrared back-reaction will be physically observable. The
situation will be very much analogous to what happens in the case of
parametric resonance of gravitational fluctuations during inflationary
reheating. This process is a gauge artifact in single field models of
inflation \cite{Fabio1} (see also \cite{Weinberg2,Parry,Zhang}), 
but it is real and unsuppressed in certain
two field models \cite{BV,Fabio2}. In the case of two field
models, work on the analysis of
the back-reaction effects of infrared modes on the observable representing
the local Hubble expansion rate is in progress. 

Provided that it can indeed be shown that infrared modes have a
nontrivial gravitational back-reaction effect in interesting models
at second order in perturbation theory, it then becomes important
to extend the analysis beyond perturbation theory. For initial attempts
in this direction see \cite{WT3,AWT}.

\vskip 0.5cm

\centerline{Acknowledgements}

I wish to thank the organizers of the Vth Mexican School 
for inviting me to lecture at
this wonderful place on the coast of the Yucatan peninsula. I am grateful
to my collaborators Raul Abramo, Fabio Finelli, Ghazal Geshnizjani, 
Pei-Ming Ho, Sergio Joras, Jer\^ome Martin and in particular Slava
Mukhanov for sharing their insights. This work has been supported in
part by the U.S. Department
of Energy under Contract DE-FG02-91ER40688, TASK A.
%
%
%
\input{referenc}



\printindex
\end{document}

%% file: referenc.tex
%
%

%
%